\documentclass[useAMS,usenatbib]{mn2e}
\input psfig.tex
\usepackage{amsmath}
\usepackage{graphicx}
\usepackage{rotating}		

        \newcommand{\Hbeta}{\mbox{${\rm  H_{\beta}}$}}
        \newcommand{\MgFe}{\mbox{${\rm  [MgFe]} $}}
        \newcommand{\MgFeb}{\mbox{${\rm  [MgFe]'} $}}
        \newcommand{\MFe}{\mbox{${\rm \langle Fe \rangle} $}}
        \newcommand{\mgii}{\mbox{${\rm Mg_{2}} $}}
        
        \newcommand{\mgb}{\mbox{${\rm Mg_{b}} $}}
        \newcommand{\nad}{\mbox{${\rm NaD} $}}
        \newcommand{\cii}{\mbox{${\rm C_{2}4668} $}}
        \newcommand{\hga}{\mbox{${\rm H_{\gamma A}} $}}
        \newcommand{\hda}{\mbox{${\rm H_{\delta A}} $}}
        \newcommand{\hgf}{\mbox{${\rm H_{\gamma F}} $}}
        \newcommand{\hdf}{\mbox{${\rm H_{\delta F}} $}}

	\newcommand{\FeH}{\mbox{[Fe/H]}}

	\newcommand{\asfe}{\mbox{[$\alpha$/Fe]}}
        \newcommand{\alfa}{\mbox{$\alpha$-elements}}
        \newcommand{\alfe}{\mbox{$\alpha$-enhanced}}
        \newcommand{\enh}{\mbox{$\alpha$-enhancement}}
        
    \newcommand{\Msun}{\mbox{${\rm M_{\odot}}$}}   
    \newcommand{\Xsun}{\mbox{${\rm X_{\odot}}$}}   
    \newcommand{\Zsun}{\mbox{${\rm Z_{\odot}}$}}   
    \newcommand{\Teff}{\mbox{${\rm T_{eff}}$}}  
    \newcommand{\logT}{\mbox{${\rm \log}$~${\rm T_{eff}}$}}
    \newcommand{\logG}{\mbox{${\rm \log{(g)}}$}}

        \def\oneskip{\vskip 4pt}
        \def\smallskip{\vskip 3pt}
        \def\littleskip{\vskip 2pt}

\title[Measuring age, metallicity and abundance ratios from 
      absorption line indices]{Measuring age, metallicity and abundance ratios from 
      absorption line indices}

\author[R. Tantalo \& C. Chiosi]{Rosaria Tantalo \& Cesare Chiosi\\
 Department of Astronomy, University of Padova,
       Vicolo dell'Osservatorio 2, 35122 Padova, Italy\\
E-mail: {\tt tantalo@pd.astro.it; chiosi@pd.astro.it}
}

\date{\tt Submitted: November 2003; Revised: May 2004}

\pubyear{2004}

\begin{document}
\maketitle
\title{Measuring age, metallicity and abundance ratios from 
      absorption line indices}

\begin{abstract}
In this study we present detailed calculations of absorption line
indices on the Lick System based on the new stellar models by
\citet{Salasnich20} incorporating the enhancement of \alfa\ 
both in the opacity and in the chemical abundances. The models span
large ranges of initial masses, chemical compositions, and ages, and
are calculated for both solar and enhanced abundance ratios $\rm
[X_{el}/Fe]$ of \alfa. With these models and the so-called {\it
Response Functions} of \citet{Tripicco95}, we calculate the indices
for Single Stellar Populations (SSPs) of different age, metallicity
and degree of enhancement. Starting from the widely accepted
conviction that \Hbeta\ is a good age indicator, that [MgFe] is most
sensitive to metallicity, and indices like \mgb, \mgii\ and others are
most sensitive to metallicity and degree of enhancement, we made use
of the triplet \Hbeta, \mgb\ and \MFe, and {\it Minimum-Distance
Method} proposed by \citet{Trager20a} to estimate the age, metallicity
and enhancement degree for the galaxies of the \citet{Gonzalez93}
sample, and compare the results with those by \citet{Trager20a} and
\citet{Thomas03}. Since very large differences are found, in particular 
as far as the age is concerned, ours are systematically older than
those of \citet{Trager20a} and \citet{Thomas03}, we analyze in a great
detail all possible sources of disagreement, going from the stellar
models and SSPs to many technical details of the procedure to
calculate the indices, and finally the pattern of chemical elements
(especially when \alfe\ mixtures are adopted). Each of the above
aspects of the problem bears on the final result: amazingly enough, at
increasing complexity of the underlying stellar models and SSPs, the
uncertainty increases. However, the key issue of the analysis is that
at given metallicity Z and enhancement factor, the specific abundance
ratios $\rm [X_{el}/Fe]$ adopted for some elements (e.g. O, Mg, Ti,
and likely others) dominate the scene because with the
\citet{Tripicco95} {\it Response Functions} they may strongly affect
indices like \Hbeta\ and the age in turn. In brief, with the ratio
[Ti/Fe]=0.63 adopted by \citet{Salasnich20}, \Hbeta\ at old ages
turned out to be larger than the mean observational value, and therefore
the age was forced to very old values in order to recover the
observations.  In contrast, the results by \citet{Trager20a} and
\citet{Thomas03} are immediately recovered if their [Ti/Fe] ratios are
adopted, i.e. [Ti/Fe]=0.0 or 0.3, respectively.  We have also analyzed
how the galaxy ages, metallicities and degrees of enhancement vary
with the triplets of indices in usage. To this aim we turn to the
Trager ``{\it IDS Pristine}'' sample which contains many more galaxies
and a much wider list of indices than the Gonz\'alez sample. The
solution is not unique in that reflecting the poor ability of most
indices to disentangle among the three parameters. Finally, at the
light of the above results and points of uncertainty, we have drawn
some remarks on the interpretation of the distribution of early-type
galaxies in popular two-indices planes, like \Hbeta\ vs. \MgFe.  We
argue that part of the scatter along the \Hbeta\ axis observed in this
plane could be attributed instead of the age, the current explanation,
to a spread both in the degree of enhancement and some abundance
ratios. If so, another dimension is added to the problem, i.e. the
history of star formation and chemical enrichment in individual
galaxies. The main conclusion of this study is that deriving ages,
metallicities and degree of enhancement from line indices is a
cumbersome affair whose results are still uncertain.
\end{abstract}


\section{Introduction}\label{intro}

Over the years, much effort has been spent to infer from observational
data the age, metallicity, and enhancement of the \alfa\ of the
stellar content in early-type galaxies (EGs)
\citep{Bressan96,Tantalo98,Tantalo98a,Jorgensen99,Trager20a,Trager20b,
Kuntschner00,Kuntschner01b,Vazdekis01,Davies2001,
Maraston03,Thomas03,Thomas03a,ThoMara03}.

Determining the age and metallicity is a cumbersome affair hampered by
the fact that the spectral energy distribution of an old metal-poor
stellar population may happen to be the same of a young metal-rich
one. This is otherwise known as the {\it age-metallicity degeneracy}
pointed out long ago by \citet{Renbuz86}.

A promising way-out is perhaps offered by the absorption line indices
defined by the Lick group \citep{Worthey92,Worthey94}, which seem to
have the potential of partially resolving the {\it age-metallicity
degeneracy}. An extensive use of the two indices diagnostics is
currently made in order to infer the age and the metallicity of
early-type galaxies
\citep{Kuntschner98a,Jorgensen99,Kuntschner98,Kuntschner00,Kuntschner01b,
Trager20a,Vazdekis01,Davies2001,Poggianti01,Maraston03,Thomas03,Thomas03a,
ThoMara03}.

The problem is, however, further complicated by a third parameter,
i.e. the abundance ratio \asfe\ (where $\alpha$ stands for all
chemical elements produced by $\alpha$-captures on lighter nuclei).
Absorption line indices like \mgii\ and \MFe\ measured in the central
regions of galaxies are known to vary passing from one galaxy to
another \citep{Gonzalez93,Trager20a,Trager20b}. Looking at the
correlation between \mgii\ and \MFe\ (or similar indices) for the
galaxies in the above quoted samples, \mgii\ increases faster than
\MFe, which is interpreted as due to enhancement of \alfa\ in some
galaxies. In addition to this, since the classical paper by
\citet{Burstein88}, the index \mgii\ is known to increase with the
velocity dispersion (and hence mass and luminosity) of the galaxy.
Standing on this body of data the conviction arose that the degree of
enhancement in \alfa\ ought to increase passing from dwarf to massive
EGs \citep{Faber92,Worthey94,Mattfra94,Matteucci97,Matteucci98}. The
simplest and most widely accepted interpretation is the one based on
the different duration of the star forming period and the different
contribution to \alfa\ and Fe by Type II supernov\ae\ from massive stars
(mostly producing \alfa) and Type Ia supernov\ae\ from accreting white
dwarfs in binary systems (mostly generating Fe). Since the minimum
mean time scale for a binary mass-accreting white dwarf to get the
supernova stage is $\geq$0.5\,Gyr, the iron contamination by Type Ia
supernov\ae\ occurs later as compared to the ones of Type II
\citep{Greggio83}. With the standard supernova driven galactic wind
model (SDGW) by \citet{Larson74} and classical initial mass function,
to reproduce the observed trend of the $\rm [\alpha/Fe]$-mass
relationship, the total duration of the star forming activity ought to
decrease at increasing galaxy mass. Long ago \citet{Larson74}
suggested that the Color-Magnitude Relation stems from a mass-mean
metallicity relation, whereby the massive EGs are more metal-rich. He
also proposed the SDGW mechanism as the key process by which massive
EGs, owing to their deeper gravitational potential, retain gas and
form stars for longer periods of time than the low-mass ones. {\it
This is the opposite of what implied by the trend in
$\alpha$-elements}. The kind of behavior for the star formation
history in EGs required by the $\alpha$-enhancement has been suggested
by \citet{Bressan96} and \citet{Tantalo98a} on the base of their
properties in the two indices planes, and more recently confirmed by
\citet{Poggianti01} and the N-body-Tree-SPH models of EGs by
\citet{Chiocar02}.

In presence of \alfe\ chemical compositions, ages and metallicities of
EGs should be derived from indices in which this effect is taken into
account. As long ago noticed by \citet{Worthey92b} and \citet{Weiss95}, 
indices for \alfe\ chemical mixtures of given total metallicity are expected 
to differ from those of the standard case.

The first attempt to simultaneously derive ages, metallicities and
\asfe\ ratios for the early-type galaxies of the \citet{Gonzalez93} 
sample is by \citet{Tantalo98a} by means of the so-called
$\Delta$-Method in which the effect of different ratios \asfe\ on 
theoretical indices has been included assuming the \citet{Borges95} 
calibration for \mgii\ and a suitable relationship between total 
metallicity $Z$ and iron content \FeH\ \citep{Bressan96,Tantalo98a}. 
Similar attempt was made by \citet[][TFWG00]{Trager20a} using different 
calibrations \citep{Tripicco95} but reaching similar results.

However, the SSPs adopted by \citet{Bressan96}, \citet{Tantalo98},
\citet{Tantalo98a} and also \citet{Tantalo98b} were actually
calculated from the Padova library of stellar models with solar
partition of elements and old opacities \citep{Bertelli94}.
The SSPs adopted by TFWG00 are taken from the even older SSPs library by
\citet{Worthey94a}. Recently \citet{Thomas03}, \citet[][TMB03]{Thomas03a}, 
\citet{Maraston03} and \citet{ThoMara03} presented synthetic  
absorption line indices on the Lick system for SSPs with variable
chemical abundances calibrated on a sample of Milky Way Globular
Clusters whose metallicities vary from ${\rm Z_{\odot}/30}$ to
${\rm Z_{\odot}}$. In general almost all of the above studies are based on
stellar models whose chemical composition (in particular the abundance
ratios) are not the same as those used to derive the indices. Stellar
models and indices are in a sense decoupled. Since in the meantime
more recent stellar models, isochrones, and SSPs with updated physical
input and \alfe\ chemical mixtures have become available
\citep[][SGWC00]{Salasnich20}, in the present study we intend to
bridge the gap and to generate and use indices that are fully
compatible with the stellar models underneath, and cast light on how
much the results would depend on the internal consistency for the
chemical parameters.
 
The plan of the paper is as follows. Section~\ref{inddef} defines the
so-called {\it enhancement factor} for a mixture in which the
abundance of \alfa\ with respect to that of Fe is enhanced as compared
to the Solar Mix. Section~\ref{newiso} presents a brief description of
the new stellar models and SSPs by SGWC00 in which \alfe\ chemical
abundances are adopted. Section~\ref{def_ind} shortly introduces the
definition of the absorption line indices for a single star, and their
corresponding integrated values for SSPs. Section~\ref{calibr}
presents three different ways or ``calibrations'' to include the
effect of \alfe\ chemical compositions on absorption line indices and
the choice we have made. Section~\ref{build_ind} describes the
formalism we have used to derive the integrated indices for SSPs.
Section~\ref{method} firstly presents the indices for SSPs of different
age, metallicity and enhancement factor, secondly illustrates the method
to derive these parameters from the observational indices, contains
some preliminary results for a selected sample of EGs, and finally
compares them with those from previous studies. Since very large
differences are found, in Section~\ref{why} we analyze in detail the
many reasons why such large differences may occur, i.e. the stellar
models in use, the coverage of evolutionary phases, and other
technical aspects in the calculations of SSP indices.
Section~\ref{threecase} is dedicated to the specific subject of the
effects caused by using different pattern of abundances for
$\alpha$-enhanced elements, pointing out that this is the parameter
driving the whole subject. Specifically we investigate the role played
by some important elements, and argue that the source of the
disagreement between the results obtained by TFWG00, \citet{Thomas03} 
and the ones presented here is entirely due to the different abundance 
ratio [Ti/Fe] adopted by SGWC00 (Section~\ref{Titanium}). In
Section~\ref{unique} we examine whether the solution for the age,
metallicity and enhancement factor is unique, in the sense that the
same result is obtained using different combinations of indices in
groups of three. As expected no unique answer is found. To overcome
the difficulty we check whether imposing the simultaneous fit of many
indices the solution can be better constrained, In
Section~\ref{twoind} we draw some remarks on the widely adopted
two-indices plane diagnostics to estimate the age, metallicity, and
degree of enhancement of \alfa. Finally, a summary of the results of
this study and some concluding remarks are presented in
Section~\ref{concl}.

\section{Chemical enhancement: definition }\label{inddef}

In presence of enhancement in \alfa\ one has to modify the
relationship between the total metallicity Z and the iron content
\FeH. This is made by suitably defining the enhancement parameter
$\Gamma$. Following \citet{Tantalo98a}, let us split the metallicity
Z in the sum of two terms

\begin{equation}
{\rm Z = \sum_{j} X_{j} + X_{Fe}}
\label{sum1}
\end{equation}

\noindent
where ${\rm X_j}$ are the abundances by mass of all heavy elements but Fe, 
and ${\rm X_{Fe}}$ is the same for Fe. Recasting eqn.~(\ref{sum1}) as

\begin{equation}
{\rm Z = \frac{X_{Fe}}{X_H} X_H \left[ 1+ \frac{\sum_{j} X_{j}}{X_{Fe}} \right]}
\label{sum2}
\end{equation}

\noindent
and normalizing it to the solar values we get

\begin{equation}
{\rm 
\left(\frac{Z}{Z_{\odot}}\right)= 
           \left(\frac{X_{Fe}}{X_{H}} \right) 
           \left(\frac{X_{H}}{X_{Fe}} \right)_{\odot}
           \left(\frac{X_{H}}{X_{H,\odot}} \right) 
           \frac{\left[1+ \frac{\sum_{j} X_{j}}{X_{Fe}}\right]~}
           {\left[1+ \frac{\sum_{j} X_{j}}{X_{Fe}}\right]_{\odot}}
}
\label{sum3}
\end{equation}

\noindent 
from which we finally obtain  

\begin{equation}
{\rm [Fe/H] = \log{(Z/\Zsun)} - \log{(X/\Xsun)} - \Gamma}
\label{feh1}
\end{equation}

\noindent
For solar metallicity \Zsun\ and solar-scaled mixture
($\Gamma$=0), eqn.~(\ref{feh1}) yields \FeH=0. The definition of
$\Gamma$ is obvious. There in after $\Gamma$ is referred to as the
{\it total enhancement parameter}. This relation is useful to
re-scale the Fe content in presence of $\Gamma\neq$0 and vice-versa.

To avoid misunderstanding it may be useful to briefly describe the
practical derivation of $\Gamma$ for a set of elements, whose total
metal abundance is Z, but in which a sub-group have abundances
enhanced with respect of the corresponding solar value. Let denote
with ${\rm N_{j}}$ the number density of the generic element j with
mass abundance ${\rm X_{j}}$ and define the quantity ${\rm A_{j}}$ as
in \citet{Grevesse96}

\begin{equation}
{\rm A_j = log \left(\frac{N_{j}}{N_{H}} \right) + 12}
\label{eqn5}
\end{equation}

\noindent 
The abundance by mass with respect to iron is given by

\begin{equation}
{\rm 
\left[\frac{X_{j}}{X_{Fe}} \right] = log \left( \frac{X_{j}}{X_{Fe}} \right) 
- log \left( \frac{X_{j}}{X_{Fe}} \right)_{\odot}
}
\end{equation}

\noindent 
which corresponds to

\begin{equation}
{\rm 
\left[\frac{X_{j}}{X_{Fe}} \right] = 
\left[\ \frac{X_{j}}{X_{H}} \right] - \left[\ \frac{X_{Fe}}{X_{H}} \right]
}
\end{equation}

\noindent
This can be rewritten as function of the number density ${\rm N_{j}}$ in the following way

\begin{equation}
{\rm 
\left[\frac{X_{j}}{X_{Fe}} \right] = 
\left[\ \frac{N_{j}}{N_{H}} \right] - \left[\ \frac{N_{Fe}}{N_{H}} \right]
}
\end{equation}

\noindent 
or in terms of ${\rm A_{j}}$

\begin{equation}
{\rm 
\left[\frac{X_{j}}{X_{Fe}} \right] = \left( A_{j}^{{\it enh}} - A_{j}^{\odot} \right) - 
\left( A_{Fe}^{{\it enh}} - A_{Fe}^{\odot} \right)
}
\end{equation}

\noindent
Keeping constant the number density of iron, i.e. 
${\rm (A_{Fe}^{{\it enh}}- A_{Fe}^{\odot})}$=0, we get

\begin{equation}
{\rm 
A_{j}^{{\it enh}} = \left[\frac{X_{j}}{X_{Fe}} \right] + A_{j}^{\odot}
}
\end{equation}

\noindent
Using ${\rm [X_{j}/X_{Fe}]}$ and ${\rm A_{j}^{\odot}}$ taken from
observational data and \citet{Grevesse96}, respectively, we can
calculate the new mass abundances for the solar-scaled case by means
of eqn.~(\ref{eqn5}). In general, the summation of the new mass
abundances in the set of enhanced elements will be different from the
assigned total metallicity. The mass abundances must be rescaled to
the true value ${\rm X_{j}'}$ by means of the relation

\begin{equation}
{\rm 
X'_{j}  = \frac{X_{j}}{\sum X_{j}}
}
\end{equation}

\noindent
Finally the total enhancement factor $\Gamma$ is simply given by

\begin{equation}
{\rm 
\Gamma = - log \left( \frac{X'_{Fe}}{X_{Fe}^{\odot}} \right)
}
\label{def2gam}
\end{equation}

\noindent 
which is another way of defining the enhancement factor fully
equivalent to that of eqn.~(\ref{feh1}).

A slightly different definition of enhancement has been adopted by
TFWG00 and TMB03 which however are fully equivalent to ours so that
straight comparison is possible.

Finally, it is worth calling attention that at given total metallicity
Z different patterns of ${\rm [X_j/X_{Fe}]}$ may yield the same total
enhancement factor $\Gamma$. This fact bears very much on the
correction of indices for enhancement because each elemental species
brings a different effect (see Section~\ref{cal_trip} below).

\section{Stellar models with $\alpha$-enhanced chemical compositions}\label{newiso}

SGWC00 have presented four sets of stellar models with different
initial chemical compositions [Y=0.250, Z=0.008], [Y=0.273, Z=0.019],
[Y=0.320, Z=0.040] and [Y=0.390, Z=0.070], initial masses from 0.15 to
20\, \Msun, helium-to-metal enrichment law ${\rm Y=Y_{p}+2.25 Z}$
(${\rm Y_{p}}$=0.23 is the primordial helium abundance), and both
solar and \alfe\ partitions of chemical elements.

\noindent
In Table~\ref{enh-deg} we list the usual parameters X, Y and Z
defining the initial chemical composition of a star and the
corresponding [Fe/H] ratio for solar-scaled and \alfe\ mixtures.
Columns (1) through (3) give the initial chemical composition of the
adopted stellar models, whereas columns (4), (5) and (6) list the
values of \FeH\ for the standard solar-scaled composition
($\Gamma$=0), the \alfe\ mixture ($\Gamma$=0.35) adopted by
SGWC00\footnote{It is worth noticing that SGWC00 have enhanced the
\alfa\ (see their Table 1 and Table~\ref{tab-enh} in this work) in
such a way that the iron abundances for solar metallicity is equal to
\FeH=--0.3557 or equivalently $\Gamma\simeq 0.35$}, and another choice
for an \alfe\ composition ($\Gamma$=0.50) calculated by us for the
purposes of this study and to be discussed below.

\begin{table}
\normalsize
\begin{center}
\caption[]{The \FeH\ ratio for solar-scaled and \alfe\ mixtures as a function of
 the initial chemical composition [X,Y,Z].}
\label{enh-deg}
\small
\begin{tabular}{|c c c | r| r| r|}
\hline
\multicolumn{1}{|c}{} &
\multicolumn{1}{c}{} &
\multicolumn{1}{c|}{} &
\multicolumn{1}{c|}{$\Gamma=0$} &
\multicolumn{1}{c|}{$\Gamma=0.35$} & 
\multicolumn{1}{c|}{$\Gamma=0.50$}\\
\hline
\multicolumn{1}{|c}{Z} &
\multicolumn{1}{c}{Y} &
\multicolumn{1}{c|}{X} &
\multicolumn{1}{c|}{[Fe/H]} &
\multicolumn{1}{c|}{[Fe/H]} &
\multicolumn{1}{c|}{[Fe/H]} \\
\hline
 0.008& 0.248 & 0.7440 & --0.3972 & --0.7529 &--0.8972\\
 0.019& 0.273 & 0.7080 &   0.0000 & --0.3557 &--0.5000\\
 0.040& 0.320 & 0.6400 &   0.3672 &   0.0115 &--0.1328\\
 0.070& 0.338 & 0.5430 &   0.6824 &   0.3267 &  0.1715\\
\hline
\end{tabular}
\end{center}
\end{table}

\noindent
Table~\ref{tab-enh} lists the detailed pattern of elements and
abundance ratios adopted by SGWC00 for the solar-scaled compositions
($\Gamma$=0), the \alfe\ mixtures with $\Gamma$=0.35, and our case
with $\Gamma$=0.50. Columns (2), (4) and (8) show the abundance ${\rm
A_{el}}$ of elements in logarithmic scale, columns (3), (5) and (9)
display the ratio ${\rm X_{el}/Z}$, columns (6) and (10) list the
ratio ${\rm [X_{el}/Fe]}$, and finally columns (7) and (11) show the
ratio ${\rm [X_{el}/H]}$.

The enhancement factors for individual species come from the
determinations of chemical abundances in metal-poor field stars by
\citet{Ryan91}. We call attention on the high enhancement factor for
Ti, which amounts to [Ti/Fe]=0.63. The analysis below will clarify
that the abundance of Ti bears very much on the final result, in
particular as far as the \Hbeta\ is concerned. More recent
determinations by \citet{Carney96} and \citet{Habgood01} of the
[Ti/Fe] in globular clusters yield ${\rm \langle [Ti/Fe]
\rangle\simeq}$0.25-0.30. A similar study by \citet{Gratton03} of
abundances for a sample of metal-poor stars (150 objects in total)
with accurate parallaxes yields ${\rm
\langle[Ti/Fe]\rangle\simeq}$0.20$\pm$0.05 with long tails on both
sides, the highest values however rarely exceeding 0.4. In spite of
this, since the SGWC00 models have been calculated with [Ti/Fe]=0.63,
we adopt this value throughout our study to compare the indices and
their follow up with those of similar studies. Only at very end we
will discuss the consequences of choosing different (lower)
enhancement factors for this element.

The stellar models calculated by SGWC00 for $\Gamma$=0 and
$\Gamma$=0.35 extend from the ZAMS up to either the start of the
thermally pulsing asymptotic giant branch (TP-AGB) phase or carbon
ignition. The major novelty of these stellar models with respect to
previous calculations \citep[e.g][]{Bertelli94,Girardi96} is the
opacity whose chemical mixture is the same of the stellar models. No
details on the stellar models are given here; they can be found in
SGWC00. Suffice it to mention that: (i) in low mass stars passing
from the tip of red giant branch (T-RGB) to the HB or clump, mass-loss
by stellar winds is included according to the \citet{Reimers75} rate
with $\eta$=0.45; (ii) the whole TP-AGB phase is included in the
isochrones with ages older than 0.1\,Gyr according to the algorithm of
\citet{GirBer98} and \citet{Girardi20}. Because of the opacities used
by SGWC00 the new tracks of high metallicity (Z=0.070) do not
develop the so-called hot horizontal branch (H-HB) and AGB-manqu\`e
phase for low mass stars \citep[see][ for all details]{Greggio90,Bressan94} 
thus affecting a potential source of energy in the UV and visible ranges of 
the spectra with some important consequences for indices like \Hbeta, 
\hgf, etc. Finally for the four sets of stellar models the corresponding 
isochrones and SSPs are calculated\footnote{The complete grids of
stellar tracks, isochrones and SSPs are available on the web site {\it
http://pleiadi.pd.astro.it}.}.

\begin{table*}
\normalsize
\begin{center}
\caption[]{Abundance ratios for the solar-scaled and \alfe\ mixtures adopted in this study. 
The case for $\Gamma=0.35$ is the same as in SGWC00, the case of
$\Gamma=0.50$ has been extrapolated by us (see the text for details).}
\label{tab-enh}
\small
\begin{tabular*}{154mm}{|c| c c | c c c c|  c c c c|}
\hline
\multicolumn{1}{|c|}{} &
\multicolumn{2}{c|}{$\Gamma=0$} &
\multicolumn{4}{c|}{$\Gamma=0.35$} &
\multicolumn{4}{c|}{$\Gamma=0.50$}\\
\hline
\multicolumn{1}{|c|}{Element} &
\multicolumn{1}{c}{${\rm A_{el}}$} &
\multicolumn{1}{c|}{${\rm X_{el}/Z}$} &
\multicolumn{1}{c}{${\rm A_{el}}$} &
\multicolumn{1}{c}{${\rm X_{el}/Z}$} &
\multicolumn{1}{c}{${\rm [X_{el}/Fe]}$} &
\multicolumn{1}{c|}{${\rm [X_{el}/H]}$} &
\multicolumn{1}{c}{${\rm A_{el}}$} &
\multicolumn{1}{c}{${\rm X_{el}/Z}$} &
\multicolumn{1}{c}{${\rm [X_{el}/Fe]}$} &
\multicolumn{1}{c|}{${\rm [X_{el}/H]}$} \\
\hline
  O   & 8.870 & 0.482273 & 9.370 & 0.672836 & 0.50 &  +0.1442 &  9.570 & 0.770549 & 0.70 &  +0.2035 \\
  Ne  & 8.080 & 0.098668 & 8.370 & 0.084869 & 0.29 & --0.0658 &  8.490 & 0.079983 & 0.41 & --0.0912 \\
  Mg  & 7.580 & 0.037573 & 7.980 & 0.041639 & 0.40 &  +0.0441 &  8.140 & 0.043451 & 0.56 &  +0.0631 \\
  Si  & 7.550 & 0.040520 & 7.850 & 0.035669 & 0.30 & --0.0558 &  7.970 & 0.033806 & 0.42 & --0.0787 \\
  S   & 7.210 & 0.021142 & 7.540 & 0.019942 & 0.33 & --0.0258 &  7.670 & 0.019498 & 0.46 & --0.0352 \\
  Ca  & 6.360 & 0.003734 & 6.860 & 0.005209 & 0.50 &  +0.1441 &  7.060 & 0.005970 & 0.70 &  +0.2038 \\
  Ti  & 5.020 & 0.000211 & 5.650 & 0.000387 & 0.63 &  +0.2634 &  5.890 & 0.000495 & 0.87 &  +0.3703 \\
  Ni  & 6.250 & 0.004459 & 6.270 & 0.002056 & 0.02 & --0.3371 &  6.280 & 0.001503 & 0.03 & --0.4723 \\
  C   & 8.550 & 0.173285 & 8.550 & 0.076451 & 0.00 & --0.3553 &  8.550 & 0.054828 & 0.00 & --0.4998 \\
  N   & 7.970 & 0.053152 & 7.970 & 0.023450 & 0.00 & --0.3553 &  7.970 & 0.016827 & 0.00 & --0.4995 \\
  Na  & 6.330 & 0.001999 & 6.330 & 0.000882 & 0.00 & --0.3553 &  6.330 & 0.000632 & 0.00 & --0.5001 \\
  Cr  & 5.670 & 0.001005 & 5.670 & 0.000443 & 0.00 & --0.3557 &  5.670 & 0.000318 & 0.00 & --0.4997 \\
  Fe  & 7.500 & 0.071794 & 7.500 & 0.031675 & 0.00 & --0.3557 &  7.500 & 0.022751 & 0.00 & --0.4991 \\
\hline
\end{tabular*}
\end{center}
\end{table*}

\section{Absorption line indices for SSPs}\label{def_ind}

Although the definition of the absorption line indices and how these
are calculated for SSPs can be found in the original papers by
\citet{Burstein84}, \citet{Faber85}, \citet{Worthey92}, \citet{Worthey94}, 
and \citet{Bressan96}, it may be useful to summarize here the various
steps of the procedure. For the sake of easy comparison, we strictly
follow the notation adopted by \citet{Maraston03}.

The definition of an absorption line index with passband
$\Delta_{\lambda}$ is different according to whether it is measured in
equivalent width (EW) or magnitude as given by

\begin{equation}
\begin{aligned}
{\rm I_{l}} &= {\rm \Delta_{\lambda} \left(1 - \frac{F_{l}}{F_{c}} \right)} ~~~~~{\rm in ~EW}\\
{\rm I_{l}} &= {\rm -2.5 \log \left(\frac{F_{l}}{F_{c}} \right)} ~~~~ {\rm in ~Mag}
\end{aligned}
\label{ind_def}
\end{equation}

\noindent
where ${\rm F_{l}}$ and ${\rm F_{c}}$ are the fluxes in the line and
pseudo-continuum, respectively. Since the Lick system of indices
\citep{Burstein84,Faber85,Worthey94} stands on a spectra library with
fixed mean resolution of 8\,\AA, whereas most of the spectral
libraries in use have a different resolution, the straightforward
application of eqns.~(\ref{ind_def}) is not possible. The difficulty
is with ${\rm F_{l}}$ which depends on the spectral resolution. To
overcome this problem, the so-called {\it Fitting Functions} have been
introduced.  They express the indices measured on the observed spectra
of a large number of stars with known gravity, \Teff, and chemical
composition as functions of these parameters \citep{Worthey94}.

Given these premises, the integrated indices of SSPs can be derived in
the following way. We start from the flux in the absorption line of
the generic i-th star of the SSP, ${\rm F^{*}_{l,i}}$

\begin{equation}
\begin{aligned}
{\rm F^{*}_{l,i}} &= {\rm F^{*}_{c,i} \left( 1 - \frac{I^{*}_{l,i}}
               {\Delta_{\lambda}} \right)} ~~~~{\rm in ~EW} \\
{\rm F^{*}_{l,i}} &= {\rm F^{*}_{c,i} 10^{-0.4 I^{*}_{l,i}}} ~~~~~~~~{\rm in ~Mag}
\end{aligned}
\label{eqn11}
\end{equation}

\noindent
where ${\rm I^{*}_{l,i}}$ is the index of the i-th star computed
inserting in the {\it Fitting Functions} the values of \Teff, gravity,
and chemical composition of the star, ${\rm F^{*}_{c,i}}$ is the
pseudo-continuum flux, and ${\rm F^{*}_{l,i}}$ is the flux in the
passband. The flux ${\rm F^{*}_{c,i}}$ is calculated by interpolating to
the central wavelength of the absorption line, the fluxes in the
midpoints of the red and blue pseudo-continuum bracketing the line
\citep{Worthey94}.

Known the index for a single star, we weight its contribution to the
integrated value on the relative number of stars of the same type.
Therefore the integrated index is given by

\begin{equation}
\begin{aligned}
{\rm I_{l}^{SSP}} &= {\rm \Delta_{\lambda} 
               \left( 1 - \frac{\sum_{i} F^{*}_{l,i} N_{i}}
               {\sum_{i} F^{*}_{c,i} N_{i}} \right)} ~~~~~{\rm in ~EW} \\
{\rm I_{l}^{SSP}} &= {\rm -2.5 \log \left( \frac{\sum_{i} F^{*}_{l,i} N_{i}}
               {\sum_{i} F^{*}_{c,i} N_{i}} \right)} ~~~~{\rm in ~Mag}
\end{aligned}
\label{int_ssp}
\end{equation}

\noindent
where ${\rm N_{i}}$ is the number of stars of type i-th.

When computing actual SSPs, single stars are identified to the
isochrone elemental bins defined in such a way that all relevant
quantities, i.e. luminosity, \Teff, gravity, and mass vary by small
amounts. In particular, the number of stars per isochrone bin is given
by

\begin{equation}
{\rm 
N_{i}=\int_{m_{a}}^{m_{b}}\phi(m)dm
}
\end{equation}

\noindent
where ${\rm m_{a}}$ and ${\rm m_{b}}$ are the minimum and maximum star
mass in the bin and ${\rm \phi(m)}$ is the mass function in number.

Indicating ${\rm F^{*}_{l,i} N_{i}}$ and ${\rm F^{*}_{c,i} N_{i}}$ with
$\mathcal{F}^{*}_{l,i}$ and $\mathcal {F}^{*}_{c,i}$, respectively,
the eqns.~(\ref{int_ssp}) can be written as

\begin{equation}
\begin{aligned}
{\rm {I}_{l}^{SSP}} &= {\rm \Delta_{\lambda} \left( 1 - \frac{\sum_{i} 
                 \mathcal{F}^{*}_{l,i}}{\sum_{i} 
                 \mathcal{F}^{*}_{c,i} } \right)} ~~~~~{\rm in ~EW} \\
{\rm {I}_{l}^{SSP}} &= {\rm -2.5 \log \left( \frac{\sum_{i} 
                 \mathcal{F}^{*}_{l,i}}{\sum_{i} 
                 \mathcal{F}^{*}_{c,i}} \right)} ~~~~{\rm in ~Mag}
\end{aligned}
\label{eq_ind_ssp}
\end{equation}

\noindent
These are the equations adopted to calculate the indices of SSPs.

However, in order to evaluate the contribution to the total value of
an index by stars of the same SSP but in different evolutionary
stages, it is convenient to recast eqns.~(\ref{eq_ind_ssp}) in a
slightly different way.  Naming $\mathcal{I}^{*}_{l,i}$ the mean index
of the isochrone bin, using eqns.~(\ref{eqn11}), and applying simple
algebraic manipulations, eqns.~(\ref{eq_ind_ssp}) become

\begin{equation}
\begin{aligned}
{\rm {I}_{l}^{SSP}} &= {\rm \frac{\sum_{i} \mathcal{F}^{*}_{c,i} 
                  \mathcal{I}^{*}_{l,i}}{\sum_{i} \mathcal{F}^{*}_{c,i}}} = 
                  \sum_{{\rm i}} f^{*}_{{\rm c,i}} \mathcal{I}^{*}_{{\rm l,i}} ~~~~~~~~~~{\rm in ~EW} \\
{\rm {I}_{l}^{SSP}} &= {\rm -2.5 \log \left( \frac{ \sum_{i}\mathcal{F}^{*}_{c,i}10^{-0.4 \mathcal{I}^{*}_{l,i}}}
                 {\sum_{i} \mathcal{F}^{*}_{c,i}} \right)} \\
              &= -2.5 \log \left( \sum_{{\rm i}} f^{*}_{{\rm c,i}}  
                  10^{-0.4\mathcal{I}^{*}_{{\rm l,i}}}\right) ~~~~{\rm in ~Mag} 
\end{aligned}
\label{int2_ssp}      
\end{equation}

\noindent
where $f^{*}_{{\rm c,i}} = \mathcal{F}^{*}_{{\rm c,i}} /\sum_{{\rm i}}
\mathcal{F}^{*}_{{\rm c,i}}$ (or $f^{*}_{{\rm c,i}} = \mathcal{F}^{*}_{{\rm c,i}}
/\mathcal{F}^{{\rm SSP}}_{{\rm c,i}}$) is the contribution of the
stars in each isochrone bin to the total pseudo-continuum flux of the
SSP. The same notation of eqns.~(\ref{int2_ssp}) can be used to
indicate the $j$-th evolutionary phase and to evaluate the
contribution of this to the building up of a generic index (see
Section~\ref{worvspd} below).

\section{ $\alpha$-enhanced Indices}\label{calibr}

The {\it Fitting Functions} are one of the most crucial issues of the
whole problem because it is long known that a great deal of the final
results actually depend on them. In the following we shortly summarize
the most popular ones and then make our final choice.

\subsection{The Worthey Fitting Functions}\label{cal_worth}

The widely used {\it Fitting Functions} by \citet{Worthey94}
\citep[see also][]{Ottaviani97} depend only on \Teff, gravity, and
\FeH. Therefore the effect of an \alfe\ chemical composition will act
only via the decrease in \FeH\ given by eqn.~(\ref{feh1}) whereas the
{\it Fitting Functions} themselves will be insensitive to the
abundances of the enhanced elements but for the rescaling of [Fe/H].
Therefore, they provide only a minimum effect of an \alfe\ composition.

In contrast, many studies have emphasized that absorption line indices
should also depend on the detailed pattern of chemical abundances
\citep{Barbuy94,Idiart95,Weiss95,Borges95}. To cope with this problem
two ways out are available: the empirical {\it Fitting Functions} by
\citet{Borges95} and the semi-theoretical {\it Response Functions} 
by \citet{Tripicco95}. Unfortunately, the {\it Fitting Functions} by
\citet{Borges95} are limited to \mgii\ and \nad. For all the other
indices one has to make use of the \citet{Worthey94} {\it Fitting
Functions}. Since they are insensitive to \alfe\ but for the different
[Fe/H], a great deal of the potential effect of enhancing the \alfa\
is lost.

\begin{table*}
\normalsize
\begin{center}
\caption[]{Fractional variations $\Delta I / I$ of the indices of the 
milestone stars of the TB95 calibration for the \alfe\ mixtures
adopted in this study.}
\label{ind-cor}
\small
\begin{tabular*}{130.5mm}{|r l| r|  r| r| r| r| r|}
\hline
\multicolumn{2}{|c|}{Indices}&
\multicolumn{3}{c|}{$\Gamma=0.35$}
&\multicolumn{3}{c|}{$\Gamma=0.50$}\\
\hline
\multicolumn{2}{|c|}{} &
\multicolumn{1}{c|}{Cool-Dwarf} &
\multicolumn{1}{c|}{Cool-Giants} &
\multicolumn{1}{c|}{Turn-Off}  & 
\multicolumn{1}{c|}{Cool-Dwarf} &
\multicolumn{1}{c|}{Cool-Giants} &
\multicolumn{1}{c|}{Turn-Off}  \\
\hline
 1  & ${\rm CN_{1}}$    & --0.86877 & --0.71770 &   0.06423 & --0.94202 & --0.83719 &   0.09355 \\
 2  & ${\rm CN_{2}}$    & --0.70699 & --0.64130 &   0.12677 & --0.83117 & --0.77143 &   0.18661 \\
 3  & ${\rm Ca4227}$    &   0.49139 &   1.16619 &   0.29979 &   0.75632 &   1.96846 &   0.42250 \\
 4  & ${\rm G4300}$     & --0.21532 & --0.22310 & --0.21811 & --0.30403 & --0.30946 & --0.30218 \\
 5  & ${\rm Fe4383}$    & --0.30258 & --0.23803 & --0.37851 & --0.40260 & --0.32100 & --0.50678 \\
 6  & ${\rm Ca4455}$    &   0.10268 &   0.21885 & --0.16305 &   0.15128 &   0.32776 & --0.22517 \\
 7  & ${\rm Fe4531}$    &   0.11746 &   0.00750 & --0.08348 &   0.15334 & --0.00689 & --0.13409 \\
 8  & ${\rm C_{2}4668}$ & --0.86323 & --0.74107 & --0.75182 & --0.94322 & --0.85574 & --0.87089 \\
 9  & ${\rm \Hbeta}$    &   3.84438 &     -~~~~ &   0.02082 &   7.23898 &     -~~~~ &   0.02685 \\
10  & ${\rm Fe5015}$    & --0.00723 & --0.03527 & --0.00808 & --0.02139 & --0.05729 & --0.00496 \\
11  & ${\rm Mg_{1}}$    & --0.05511 & --0.40093 & --0.39931 & --0.10092 & --0.53182 & --0.52801 \\
12  & ${\rm Mg_{2}}$    &   0.01540 & --0.00348 & --0.00323 &   0.00611 & --0.02029 & --0.02536 \\
13  & ${\rm Mg_{b}}$    &   0.17517 &   0.89365 &   0.32348 &   0.24424 &   1.48472 &   0.49262 \\
14  & ${\rm Fe5270}$    & --0.13515 & --0.20974 & --0.08969 & --0.19210 & --0.28893 & --0.15437 \\
15  & ${\rm Fe5335}$    & --0.18802 & --0.11060 & --0.26476 & --0.26154 & --0.15684 & --0.37135 \\
16  & ${\rm Fe5406}$    & --0.21219 & --0.20857 & --0.41037 & --0.29123 & --0.28739 & --0.55194 \\
17  & ${\rm Fe5709}$    &   0.07110 &   0.15444 & --0.14799 &   0.09688 &   0.22449 & --0.22470 \\
18  & ${\rm Fe5782}$    & --0.21869 & --0.17021 & --0.59314 & --0.30428 & --0.23570 & --0.74622 \\
19  & ${\rm NaD}$       & --0.18801 & --0.29678 & --0.49855 & --0.26211 & --0.40124 & --0.65122 \\
20  & ${\rm TiO_{1}}$   &   0.00000 & --0.76789 &   0.01280 &   0.00000 & --0.87503 &   0.11905 \\
21  & ${\rm TiO_{2}}$   & --0.33456 & --0.55618 & --0.26927 & --0.44872 & --0.69355 & --0.36565 \\
\hline
\end{tabular*}
\end{center}
\end{table*}

\subsection{The \citet{Tripicco95} method}\label{cal_trip} 

A method designed to include the effects of enhancement on all indices
at once has been suggested by \citet[TB95]{Tripicco95}, who introduce
the concept of {\it Response Functions}. In brief from model
atmospheres and spectra for three proto-type stars, i.e. a Cool-Dwarf
star (CD) with \Teff=4575\,K and \logG=4.6, a Turn-Off (TO) star with
\Teff=6200\,K and \logG=4.1, and a Cool-Giant (CG) star with
\Teff=4255\,K and \logG=1.9, they calculate the absolute indices ${\rm I_{0}}$. 
Doubling the abundances ${\rm X_{i}}$ of the C, N, O, Mg, Fe, Ca, Na,
Si, Cr, and Ti in steps of ${\rm \Delta [X_{i}/H]}$=0.3\,dex they
determine the incremental ratios ${\rm \Delta I_{0}/\Delta [X_{i}/H]}$
in units of the observational error $\sigma_{0}$. The absolute indices
${\rm I_{0}}$, the observational error $\sigma_{0}$, and the
normalized incremental ratios ${\rm \delta I_{0}}$ are given in Tables
4, 5 and 6 of TB95. The true incremental ratios are ${\rm \Delta
I_{0}/\Delta [X_{i}/H] 0.3=\delta I_{0} \times \sigma_{0}}$. The {\it
Response Functions} ${\rm R_{0.3}(i)}$ for any index corresponding to
the variation of the element i-th ${\rm \Delta [X_{i}/H]=+0.3}$\,dex,
is given by

\begin{displaymath}
{\rm 
R_{0.3}(i) = \frac{1}{I_{0}}\, \frac{\Delta I_{0}}{\Delta [X_{i}/H]}\, 0.3
}
\end{displaymath}
 
\noindent
The {\it Response Functions} for Cool-Dwarfs, Cool-Giants, and
Turn-Off stars constitute the milestones of the calibration. They are
used to correct the indices for solar partitions \citep{Worthey94} in
presence of $\alpha$-enhancement.

The technical methods for correcting an index from solar to
$\alpha$-enhanced element partitions have been proposed by TFWG00 and
TMB03. Since the two methods are not strictly equivalent, some
clarification is worth here.

\noindent
Without providing a formal justification, TFWG00 propose that the
fractional variation of an index to changes of the chemical parameters
is the same as that for the reference index ${\rm I_{0}}$ according to the
relation

\begin{equation}
{\rm  
{\frac{\Delta I}{I}} = {\frac{\Delta I_{0}}{I_{0}}} = 
 \left\{ \prod_{i} [1 + R_{0.3}(i)]^\frac{ [X_{i}/H]}{0.3} \right\} -1
}
\label{dind}
\end{equation}

\noindent
where ${\rm R_{0.3}(i)}$ are the {\it Response Functions} tabulated by TB95
(in view of the discussion below we call attentions on the negative
values for some ${\rm I_{0}}$ in their list). Relation~(\ref{dind}) can be
derived as follows. We start from the assumption that

\begin{equation}
{\rm {dI \over I} = const}, 
\end{equation}

\noindent
differentiate the generic index I with respect to the abundance ratios 

\begin{equation}
{\rm 
dI_{0} =  \sum_i \frac{\partial I_{0}}{\partial [X_{i}/H]} ~0.3~ \frac{d[X_{i}/H]}{0.3}
}
\end{equation}

\noindent
and write 

\begin{equation}
{\rm 
\frac{dI_{0}}{I_{0}} =  \sum_i r_{0.3}(i) \frac{d[X_{i}/H]}{0.3}
}
\label{int_ind}
\end{equation}

\noindent
where ${\rm r_{0.3}(i) = \frac{1}{I_{0}}\frac{\partial I_{0}}{\partial
[X_{i}/H]}~0.3}$. We have introduced the quantity ${\rm r_{0.3}(i)}$
containing the partial derivatives to distinguish it from the {\it
Response Functions} ${\rm R_{0.3}(i)}$ defined above. Integrating
relation~(\ref{int_ind}) we obtain

\begin{equation}
{\rm 
ln |I| = ln |I_{0}| + \sum_i r_{0.3}(i) \frac{[X_{i}/H]}{0.3}
}
\end{equation}

\noindent
which upon exponentiation becomes

\begin{equation}
{\rm 
I=  I_{0} \prod_i exp \left[ r_{0.3}(i) \right]^{\frac{[X_{i}/H]}{0.3}}
}
\label{exp}
\end{equation}

\noindent
with the obvious condition that ${\rm I/I_{0} > 0}$. Recalling that TB95
provide the ratio ${\rm \frac{\Delta I_{0}}{\Delta [X_{i}/H]}}$ evaluated
for ${\rm \Delta [X_{i}/H]}$=0.3, for each chemical element we may write

\begin{equation}
{\rm 
\left( \frac{1}{I_{0}} \frac{\Delta I_{0}}{\Delta [X_{i}/H]} \right)_{i} = 
\frac{1}{I_{0}} \frac{I_{0} exp \left[ r_{0.3}(i) \right] - I_{0}}{0.3}
}
\end{equation}

\noindent
from which we derive

\begin{equation}
{\rm 
exp \left[ r_{0.3}(i) \right] = 
      1 + \left( \frac{1}{I_{0}}~\frac{\Delta I_{0}}{\Delta [X_{i}/H]}~0.3 \right)_{i}
    = 1 + R_{0.3}(i)
}
\label{eq_exp}
\end{equation}

\noindent
Substituting the ${\rm exp \left[ r_{0.3}(i) \right]}$ into
relation~(\ref{exp}) and performing trivial algebraic manipulations we
obtain the relation~(\ref{dind}) of TFWG00, with the only condition
that ${\rm R_{0.3}(i) > -1}$. The advantage of this formulation is
that no particular constraint is required on the sign of ${\rm I_{0}}$
and the incremental ratios given by TB95 are straightforwardly
used. As a final remark, relation~(\ref{dind}) implies that the per
cent change is constant for each step of 0.3\,dex. However, as
emphasized by TFWG00, eqn.~(\ref{dind}) while securing that the
indices tend to zero for small abundances it let them increase with
the exponent ${\rm [X_{i}/H]/0.3}$. For abundances higher than ${\rm
[X_{i}/H]=+0.6}$\,dex, the exponent may became too large and
consequently the correction may diverge.

A different reasoning has been followed by TMB03. In brief, they start
from the observational hint that in Galactic stars \mgii\,${\rm
\propto exp([Mg/H])}$ \citep[see][]{Borges95}, to assume that all
indices depend exponentially on the abundance ratios, introduce the
variable ${\rm \ln I \propto [X_{i}/H]}$, and express the fractional
variation of an index to changes in the abundance ratios as

\begin{equation}
{\rm 
\frac{\Delta I}{I} = \left\{ \prod_{i} exp \left( R_{0.3}(i) \right) ^{\frac{ [X_{i}/H]}{0.3}} \right\} - 1
}
\label{dind_thomas}
\end{equation}

\noindent
Although the exponential term in (\ref{dind_thomas}) may look like to
that of eqn.~(\ref{exp}) actually they do not because the partial
derivatives, i.e. the term ${\rm r_{0.3}(i)}$, are replaced by the response
functions ${\rm R_{0.3}(i)}$.

As general remark, the use of eqns.~(\ref{dind}) and
(\ref{dind_thomas}) requires that the sign of the index to be
corrected is the same of the corresponding ${\rm I_{0}}$ in TB95. In
general this holds good. In principle, it may, however, happen that
the signs do not coincide. To overcome this potential difficulty,
TBM03 apply a correcting procedure forcing the negative reference
indices ${\rm I_{0}}$ of TB95 to become positive. This occurs in
particular for \Hbeta\ of Cool-Dwarf stars (and others as well). The
argument is that TB95 neglecting non-LTE effects underestimate the
true values of \Hbeta\ so that negative values found for temperatures
lower than about 4500\,K should be shifted to higher, positive values
(see Fig.12 in TB95). Although we may agree on this issue, we suspect
that any change to the values tabulated by TB95 may be risky for a
number of reasons: (i) The {\it Fitting Functions} have been derived
from a set of data that include a significant number of stars with
negative values of \Hbeta; (ii) The incremental ratios by TB95 have
been calculated for particular stars (stellar spectra) with assigned
\Teff, \logG, and ${\rm I_{0}}$. Changing the reference ${\rm I_{0}}$
while leaving unchanged the incremental ratios (partial derivatives)
may not be very safe. The ideal approach would be to repeat the TB95
analysis using different samples of data and baseline spectra; (iii)
The replacement of the partial derivatives with the TB95 incremental
ratios may be risky when dealing with exponential functions; (iv) The
large corrections required to shift the ${\rm I_{0}}$ for a number of
indices; (v) Finally, the use of ${\rm \ln I}$ as dependent variable
which requires that only positive values for ${\rm I_{0}}$ are
considered.

\begin{table*}
\normalsize
\begin{center}
\caption[]{Indices and their variations from \citet{Tripicco95} 
and \citet{Tantalo04c} using the new library of high resolution spectra by
\citet{Munari04}. ${\rm I_{0}}$ always denotes the value of the index for 
the solar-scaled compositions, whereas ${\rm I_{(M/H)}}$ is the value
obtained by increasing the $\alpha$-elements by +0.3\,dex in
\citet{Tripicco95} and +0.4\,dex in \citet{Tantalo04c}. The selected 
stars are the same as in \citet{Tripicco95}, whereas in the case
indicated as ``1\,\AA\ resolution spectra'' these are the closest
matches in \Teff\ and \logG\ in the library to our disposal. The
metallicity is always solar. See the text for more details.}
\label{variations}
\begin{tabular*}{102.5mm}{|c| c| r| r| r| r| r| r|}
\hline
\hline
\multicolumn{2}{|c|}{} &
\multicolumn{3}{|c|}{\citet{Tripicco95}} &
\multicolumn{3}{|c|}{1\,\AA\ Resolution Spectra}\\
\hline
\multicolumn{1}{|c|}{Index} &
\multicolumn{1}{|c|}{} &
\multicolumn{1}{|c|}{CD} &
\multicolumn{1}{|c|}{TO} &
\multicolumn{1}{|c|}{CG} &
\multicolumn{1}{|c|}{CD} &
\multicolumn{1}{|c|}{TO} &
\multicolumn{1}{|c|}{CG} \\
\hline
        &  \Teff                 & 4575   & 6200  & 4255   & 4500  & 6200  & 4200   \\
        &  \logG                 & ~~4.6  & ~~4.1 & ~~1.9  & ~~4.5 & ~~4.0 & ~~2.0  \\
\hline
        & ${\rm I_{0}}$          &  2.340 & 0.770 &  8.620 & 2.650 &  1.125 & 3.705 \\
\cii    & ${\rm I_{(M/H)}}$      &  3.044 & 1.794 & 11.180 & 4.500 &  0.800 & 4.000 \\
        & ${\rm \Delta I/I}$     &  0.300 & 1.329 &  0.297 & 0.698 &--0.289 & 0.079 \\    
\hline
        & ${\rm I_{0}}$          &--0.100 & 3.790 &  0.050 & 0.350 &  4.000 & 0.100 \\
\Hbeta  & ${\rm I_{(M/H)}}$      &--0.254 & 3.900 &--0.038 & 0.900 &  5.000 & 0.550 \\
        & ${\rm \Delta I/I}$     &  1.540 & 0.030 &--1.760 & 1.571 &  0.250 & 4.450 \\
\hline
        & ${\rm I_{0}}$          &  0.530 & 0.070 &  0.360 & 0.520 &  0.120 & 0.456 \\
\mgii   & ${\rm I_{(M/H)}}$      &  0.568 & 0.083 &  0.409 & 0.610 &  0.130 & 0.605 \\
        & ${\rm \Delta I/I}$     &  0.072 & 0.186 &  0.130 & 0.360 &  0.083 & 0.327 \\
\hline
\hline
\end{tabular*}
\end{center}
\end{table*}

In any case, to proceed further, one has to fix the abundance of those
\alfa\ that are enhanced with respect to the solar partition. With the
abundances specified in Tables~\ref{enh-deg} and \ref{tab-enh} and the
TFWG00 method (eqn.~(\ref{dind}) above) we calculate the relative
variations ${\rm \Delta I/I}$ for each index of the \citet{Worthey94}
system for the milestone stars of the TB95 calibration. The results
are summarized in Table~\ref{ind-cor} for $\Gamma$=0.35 and
$\Gamma$=0.50. To avoid misunderstanding, it is worth calling attention
that the quantities listed in Table~\ref{ind-cor} are the total
relative variations to changing all chemical abundances at once, and
not the response to changing individual species. Several remarks are
worth here: firstly adopting the TMB03 procedure the resulting ${\rm
\Delta I/I}$ are different and in some cases cannot be even
calculated; secondly comparing our ${\rm \Delta I/I}$ with those
evaluated by TMB03, in particular for some crucial elements like
\Hbeta, these latter turn out to be significantly smaller than ours
because of the increase of ${\rm I_{0}}$ at constant ${\rm \Delta
I_{0}/\Delta[X_{i}/H]}$ (see above) adopted by TMB03. This will
reflect on the nearly constant \Hbeta\ at increasing enhancement they
have found.  Finally, the fractional variation for the \Hbeta\ index
in Cool-Giant stars cannot be defined. Neither relation~(\ref{dind})
nor relation~(\ref{dind_thomas}) can be used, because the index has
changed sign with respect to ${\rm I_{0}}$. However, since in any case
the contribution from this type of stars to the total value for an SSP
will be small we simply neglect the correction.

Is there any independent argument favoring one correcting method
against the other?  Firstly, in addition to the response for separate
enhancements by 0.3 of individual species, TB95 also provide the
response to increasing all the ``metals'' at once by the same
factor. The resulting indices (limited to a few illustrative cases and
shortly indicated as ${\rm I_{([M/H])}}$) are reported in the columns
labelled ``\citet{Tripicco95}'' of Table~\ref{variations}, They should
provide a sort of upper limit to the expected correction.  Secondly,
basing on the huge library of 1\,\AA\ resolution synthetic spectra
calculated by \citet{Munari04} over a large range of \logT, \logG,
[Fe/H] and both for solar and $\alpha$-enhanced abundance ratios
(${\rm [X_{el}/Fe]}$), \citet{Tantalo04c} have derived theoretical line
absorption indices on the Lick system. The $\alpha$-enhanced mixture
is obtained by increasing at once the abundance ratios ${\rm
[X_{el}/Fe]}$ by 0.4 for O, Mg, Si, S, Ca, Ti, many of which are in
common with the present list. This is the analog of the case ``all
metals at once'' of \citet{Tripicco95}. In the following we will made
use of these results in advance of publication to evaluate ${\rm
I_{0}}$ and ${\rm I_{(M/H)}}$ and the ratio ${\rm \Delta I/I}$. The
results are presented in the tree columns of Table~\ref{variations}
labelled ``1\,\AA\ Resolution Spectra''.  Comparing the three groups
of data for ${\rm \Delta I/I}$ (i.e. those of Table~\ref{ind-cor},
those of the TB95 all metals enhanced at once, and those from the
1\,\AA\ resolution synthetic spectra) the situation is as follows: all
the three agree on \Hbeta\ and marginally for
\mgii, whereas they disagree on \cii.

\begin{figure}
\psfig{file=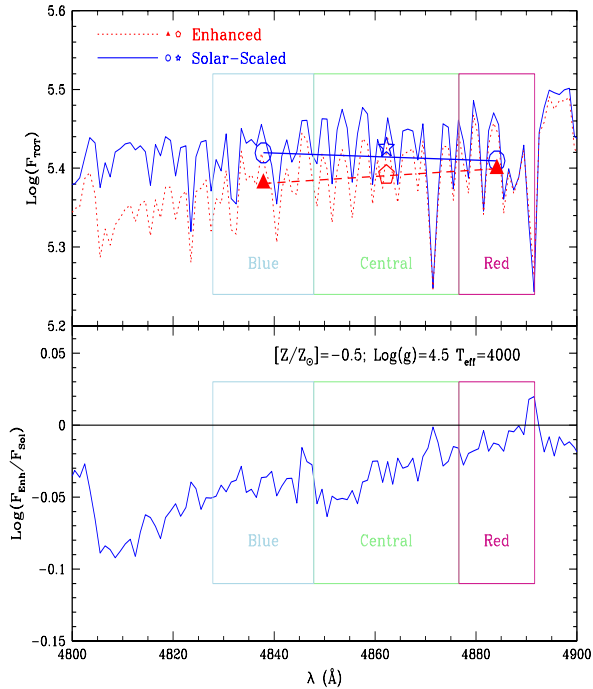,width=8.0truecm,height=10truecm} 
   \caption{Building-up of the index \Hbeta\ in the cool dwarf with
   \Teff=4000\,K, \logG=4.5, and ${\rm [Z/Z_{\odot}]}$=--0.5 both for
   the solar-scaled and the pattern of abundance ratios with ${\rm
   [\alpha/Fe]}$=+0.4. The bottom panel shows the ratio ${\rm
   F_{\lambda,enh}/F_{\lambda,\odot}}$. The absorption in the
   $\alpha$-enhanced spectrum is significantly larger than the
   solar-scaled one. The effect is larger in the blue wing of the
   pseudo-continuum and central band than in the red pseudo-continuum.
   Upper panel the spectral energy distribution of the solar-scaled
   (solid line) and $\alpha$ enhanced (dotted line) spectrum and the
   pass-bands defining the index \Hbeta. The open circles and star
   shows the mean fluxes in the three pass-band and the interpolation
   of the pseudo-continuum to derive ${\rm F_{c}}$ in the case of
   solar-scaled spectrum. The filled triangles and pentagon are the
   same but for the $\alpha$-enhanced mixture. The increase of \Hbeta\
   passing from solar to $\alpha$ enhanced abundance ratios is
   obvious.}
\label{diff_in_hb}
\end{figure}

Since a great deal of the discussion below depends on the response of
\Hbeta\ to $\Gamma$ and also to the abundance ratios adopted for some
elements at given Z and $\Gamma$, it is worth commenting on the reason
why we expect \Hbeta\ to increase with $\Gamma$ and/or ${\rm
[X_{el}/Fe]}$. The effect on other indices is less of a problem. To
this aim, with the aid of the \citet{Munari04} 1\,\AA\ resolution
spectra and the results by \citet{Tantalo04c}, limited to the case of a
typical cool-giant with \Teff=4000\,K, \logG=4.5, and ${\rm
[Z/Z_\odot]}$=--0.5 (Z=0.008), in Fig.~\ref{diff_in_hb} we show how the
index \Hbeta\ is built up and how it varies passing from ${\rm
[\alpha/Fe]}$=0.0 to ${\rm [\alpha/Fe]}$=+0.4. Firstly in the bottom panel
we display the ratio ${\rm F_{\lambda,enh}/F_{\lambda,\odot}}$ and the
pass-bands defining the index \Hbeta. The absorption in the
$\alpha$-enhanced spectrum is significantly larger than the
solar-scaled one. The effect is larger in the blue pseudo-continuum
and central band than in the red pseudo-continuum. This means that
$\alpha$-enhanced mixtures distort the spectrum in such a way that
simple predictions cannot be made. This is due to the contribution of
many molecular bands and atomic lines of elements like C, N, O, Mg,
Ti, falling in the spectral regions of interest. Secondly, in the
upper panel we show the spectral energy distribution of the
solar-scaled (solid line) and $\alpha$ enhanced (dotted line) spectrum
and once more the pass-bands for \Hbeta. The open circles and star
shows the mean fluxes in the three pass-band and the interpolation of
the pseudo-continuum to derive ${\rm F_{c}}$ in the case of solar-scaled
spectrum. The filled triangles and pentagon are the same but for the
$\alpha$-enhanced mixture. The increase of \Hbeta\ passing from solar
to $\alpha$-enhanced abundance ratios is straightforward.

Even if the ideal situation would be to replace the old {\it Response
Functions} of TB95 with new ones derived from modern high resolution
spectra, this indeed is the main target of \citet{Tantalo04c}, given
that the TFWG00 algorithm has been proved to be of general validity,
to safely use the incremental ratios of TB95, and to give results in
agreement with those from high resolution spectra, for the purposes of
the present study we prefer to use the TB95 {\it Response Functions}
combined with the TFWG00 algorithm for the sake of consistency with
results by other authors on the same subject.  Finally, it is also
easy to check that the same results of TFWG00 and TMB03 are recovered
when the same set of chemical parameters are adopted (see the Table~5
of TFWG00 and the discussion in Sect.~\ref{Titanium} below).  It is
worth recalling, however, that the final results will somewhat depend
on which correcting technique is adopted. This is a point to keep in
mind deserving further investigation.

Once the fractional variations of the milestone stars, ${\rm (\Delta
I/I)_{CD}}$, ${\rm (\Delta I/I)_{TO}}$, and ${\rm (\Delta I/I)_{CG}}$ respectively,
are known, they are linearly interpolated both in \logG\ and
\logT\ to get the total fractional variation ${\rm (\Delta I/I)}$ to be applied 
to the solar-scaled indices for each elemental bin of an isochrone
while performing the integration along it. The double interpolation
scheme assigns different weights to the calibrators according to their
distance with respect of the current elemental bin in the
HR-Diagram. The dependence of the fractional variations on gravity and
effective temperature is imposed by the TB95 study. Having only three
stars as calibrators, no other procedure can be found to account for
the change of the fractional variations along an isochrone. As a
consequence of it, as the age increases and the isochrone TO shifts to
lower \Teff\ and higher gravities, the relative weight of CD stars
(for some indices they often have the highest {\it Response
Functions}) on determining the total {\it Response Function} gets
higher. This is an additional cause of the sensitivity of indices
such as \Hbeta\ to the fractional variations.

\section{Building the indices of SSPs}\label{build_ind}

In this study we will make use only of the TB95 calibration both for
the sake of brevity and because we intend to compare our results with
those by TFWG00 who adopted the same transformation.  With the aid of
the SGWC00 SSPs, the pattern of abundances and abundance ratios
presented in Tables~\ref{enh-deg} and \ref{tab-enh}, and the TB95
calibrations and its companion corrections listed in
Table~\ref{ind-cor}, we finally derive the whole set of indices on the
Lick System for SSPs of different age, metallicity and $\Gamma$. To
apply the TB95 corrections we proceed as follows: for each elemental
bin of an isochrone and/or SSP with assigned $\Gamma$, we derive the
index according to the {\it Fitting Functions} and correct it
according to eqn.~(\ref{dind}) by interpolating the entries of
Table~\ref{ind-cor} as appropriate to the current values of \Teff, and
gravity (luminosity class), and follow the whole procedure described
in Section~\ref{def_ind}. Extensive tabulations of the complete set of
the Lick System for SSPs of different age, metallicity (Z= 0.008,
0.019, 0.040, and 0.070) and $\Gamma$ (0, 0.35 and 0.5) are
available from the authors and the web site {\it
http://dipastro.pd.astro.it/galadriel}).

\begin{figure}
\psfig{file=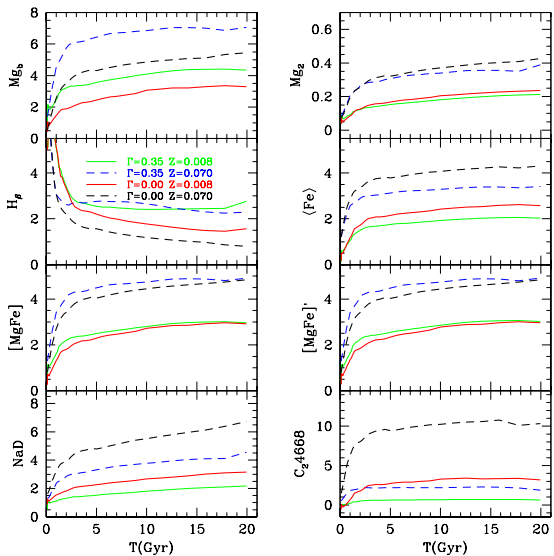,width=8.0truecm}
   \caption{Evolution of eight indices (\mgb, \mgii, \Hbeta, \MFe,
   \MgFe, \MgFeb, \nad\ and \cii) as function of the age. The heavy
   lines show the indices for solar-scaled partition of elements
   ($\Gamma$=0), whereas the thin lines show the same but for the
   partition of \alfe\ of SGWC00 ($\Gamma$=0.35). Only two
   metallicities are displayed for the sake of clarity, i.e. Z=0.008
   (solid lines) and Z=0.07 (dashed lines).}
\label{zheh}
\end{figure}

To illustrate the results we show in Fig.~\ref{zheh} the temporal
evolution of eight important indices, i.e. \mgb, \mgii, \Hbeta, \MFe,
\MgFe, \MgFeb, \nad\ and \cii\footnote{The definition of \MFe, \MgFe\ and \MgFeb\ is

\begin{displaymath}
{\rm 
\MFe =  0.5\times (Fe5270+ Fe5335)
}
\end{displaymath}

\begin{displaymath}
{\rm 
\MgFe = \sqrt{\mgb \times (0.5 \times Fe5270+0.5 \times Fe5335)}
}
\end{displaymath}

\noindent
and 

\begin{displaymath}
{\rm 
\MgFeb = \sqrt{ \mgb \times (0.72\times Fe5270+0.28 \times Fe5335)}
}
\end{displaymath}
}, for the following combinations of metallicity and
$\Gamma$, namely Z=0.008 (solid lines) and Z=0.070 (broken lines),
$\Gamma$=0 (heavy lines) and $\Gamma$=0.35 (thin lines). The age
goes from 0.01\,Gyr to 20\,Gyr. 

The merit of this set of indices is the internal consistency as far
as the chemical parameters are concerned. The stellar models and the
indices have been calculated with the same pattern of abundances.

All indices show the same behavior. For ages older than about 3\,Gyr
they all tend to flatten out ({\it age-degeneracy}). In order to
quantify the response of the indices to changes in metallicity Z,
enhancement factor $\Gamma$, and age T we calculated the relative
per cent variations

\begin{displaymath}
{\rm 
\left[ \frac{\Delta I}{I} \right]_{Z,\Gamma} \qquad
\left[ \frac{\Delta I}{I} \right]_{T,\Gamma} \qquad
 and \qquad
\left[ \frac{\Delta I}{I} \right]_{T,Z}
}
\end{displaymath}

\noindent
where I stands for the generic index, and the variations ${\rm \Delta I}$
are calculated for ${\rm \Delta Z}$=0.070--0.008=0.062, ${\rm \Delta
T}$=15--5=10\,Gyr, and $\Delta\Gamma$=0.35. The per cent
variations are evaluated at fixed values of metallicity (Z=0.008 and
0.070), total enhancement factor ($\Gamma$=0 and 0.35), and age (T=5, 10
and 15\,Gyr) as appropriate. The results are summarized in
Table~\ref{delta_percent} for the eight indices shown in Fig.~\ref{zheh}. 
These evaluations can be easily extended to any index of the
\citet{Worthey94} list. It is soon clear that \cii\ and \nad\ strongly
depend on metallicity and to a lesser extent on $\Gamma$, whereas they
scarcely depend on the age. The indices \MgFe\ and \MgFeb\ are almost
insensitive to $\Gamma$ in agreement with TMB03. All remaining indices
almost evenly depend on the three parameters with little resolving
power. Finally, there is a non negligible dependence on the total
metallicity Z for some of the indices (\Hbeta\ in particular).

\begin{table*}
\normalsize
\begin{center}
\caption[]{Relative per cent variations of the indices at changing  age, 
metallicity, and $\Gamma$. The indices are for the SSPs of SGWC00 with 
$\Gamma$=0 and $\Gamma$=0.35.}
\label{delta_percent}
\small
\begin{tabular*}{150.5mm}{|r l c r r r r r r r r|}    
\hline
\multicolumn{1}{|r}{T(Gyr)} &
\multicolumn{1}{l}{${\rm \Delta Z}$} &
\multicolumn{1}{c}{$\Gamma$} &
\multicolumn{1}{c}{$\Delta$\Hbeta} &
\multicolumn{1}{c}{}{$\Delta$\MgFe} &
\multicolumn{1}{c}{}{$\Delta$\MgFeb} &
\multicolumn{1}{c}{}{$\Delta$\MFe} &
\multicolumn{1}{c}{$\Delta$\mgb} &
\multicolumn{1}{c}{$\Delta$\mgii} &
\multicolumn{1}{c}{$\Delta$\nad} &
\multicolumn{1}{r|}{$\Delta$\cii} \\
\hline
 5.00 & 0.06 &0.00 & --29 \%  &  76 \% & 72 \% &  76 \% & 76 \% &  91 \% & 111 \% & 245 \% \\
10.00 & 0.06 &0.00 & --34 \%  &  63 \% & 60 \% &  68 \% & 58 \% &  81 \% & 107 \% & 210 \% \\
15.00 & 0.06 &0.00 & --35 \%  &  60 \% & 57 \% &  65 \% & 55 \% &  76 \% & 103 \% & 223 \% \\
 5.00 & 0.06 &0.35 &    8 \%  &  80 \% & 76 \% &  79 \% & 81 \% & 101 \% & 118 \% & 258 \% \\
10.00 & 0.06 &0.35 &   10 \%  &  69 \% & 66 \% &  71 \% & 68 \% &  88 \% & 109 \% & 225 \% \\
15.00 & 0.06 &0.35 &  --6 \%  &  63 \% & 60 \% &  66 \% & 59 \% &  74 \% &  99 \% & 225 \% \\
\hline
\multicolumn{1}{|r}{T(Gyr)} &
\multicolumn{1}{c}{Z} &
\multicolumn{1}{c}{$\Delta\Gamma$} &
\multicolumn{1}{c}{$\Delta$\Hbeta} &
\multicolumn{1}{c}{}{$\Delta$\MgFe} &
\multicolumn{1}{c}{}{$\Delta$\MgFeb} &
\multicolumn{1}{c}{}{$\Delta$\MFe} &
\multicolumn{1}{c}{$\Delta$\mgb} &
\multicolumn{1}{c}{$\Delta$\mgii} &
\multicolumn{1}{c}{$\Delta$\nad} &
\multicolumn{1}{r|}{$\Delta$\cii} \\
\hline
 5.00 & 0.008 & 0.35 & 16 \% &  9 \% &  9 \% & --19 \% & 45 \% & --10 \% &--33 \% &--78 \%  \\
10.00 & 0.008 & 0.35 & 36 \% &  3 \% &  3 \% & --21 \% & 33 \% & --12 \% &--32 \% &--79 \%  \\
15.00 & 0.008 & 0.35 & 64 \% &  3 \% &  3 \% & --21 \% & 33 \% & --10 \% &--32 \% &--79 \% \\ 
 5.00 & 0.070 & 0.35 & 76 \% & 11 \% & 11 \% & --17 \% & 49 \% &  --5 \% &--31 \% &--77 \% \\ 
10.00 & 0.070 & 0.35 &124 \% &  7 \% &  7 \% & --19 \% & 41 \% &  --8 \% &--32 \% &--79 \% \\ 
15.00 & 0.070 & 0.35 &136 \% &  4 \% &  5 \% & --20 \% & 37 \% & --11 \% &--33 \% &--79 \% \\
\hline
\multicolumn{1}{|r}{$\Delta$T(Gyr)} &
\multicolumn{1}{c}{Z} &
\multicolumn{1}{c}{$\Gamma$} &
\multicolumn{1}{c}{$\Delta$\Hbeta} &
\multicolumn{1}{c}{}{$\Delta$\MgFe} &
\multicolumn{1}{c}{}{$\Delta$\MgFeb} &
\multicolumn{1}{c}{}{$\Delta$\MFe} &
\multicolumn{1}{c}{$\Delta$\mgb} &
\multicolumn{1}{c}{$\Delta$\mgii} &
\multicolumn{1}{c}{$\Delta$\nad} &
\multicolumn{1}{r|}{$\Delta$\cii} \\
\hline
10.00 & 0.008 & 0.00 & --32 \% & 27 \% & 27 \% & 20 \% & 34 \% & 36 \%& 33 \% &  22 \% \\
10.00 & 0.070 & 0.00 & --38 \% & 16 \% & 16 \% & 13 \% & 18 \% & 25 \%& 28 \% &  14 \% \\ 
10.00 & 0.008 & 0.35 &  --4 \% & 20 \% & 20 \% & 17 \% & 24 \% & 35 \%& 35 \% &  14 \% \\
10.00 & 0.070 & 0.35 & --17 \% &  9 \% &  9 \% &  8 \% &  9 \% & 16 \%& 23 \% &   3 \% \\
\hline
\end{tabular*}
\end{center}
\end{table*}

\section{Galaxy ages, metallicities and abundance ratios}\label{method}

\subsection{Which indices to choose?}\label{suited}

As already mentioned above, the ultimate scope of a system of indices
is to derive the age, metallicity, and enhancement factor of the stars
in aggregates of different complexity going from clusters to galaxies,
the EGs in particular. To proceed further in the analysis one has to
find three or more indices, or combination of these, most sensitive to
the parameters in question. It commonly thought that indices like
\Hbeta\ are good age indicators, whereas others like \mgb, \MFe, 
\mgii\ are good metallicity indicators. However, the results presented 
in the previous sections and the discussion below will clarify that a
certain degree of degeneracy among the different parameters is always
present so that there is no clear one to one dependence of an index
from age, metallicity and degree of \enh.

Furthermore, the choice is often dictated by the available databases
of indices.  Among others, the most popular databases are the
catalogs by \citet{Gonzalez93} for nearby field early-type galaxies,
and the Trager ``{\it IDS Pristine}'' sample \citep[see][
Table~3.1]{Trager97} for the central regions of elliptical galaxies.
For the purposes of the present study, the analysis will be limited to
these samples of data, leaving aside for the time being other more
recent compilations. The \citet{Gonzalez93} catalog provides \Hbeta,
the iron-group indices, and the Mg-group indices for the region within
the radius ${\rm R_{eff}/8}$ for a sample of nearby field
galaxies. The \citet{Trager97} sample lists \Hbeta, the higher-order
Balmer line indices (\hda, \hga, \hdf, \hgf), the iron-peak indices,
\cii, and others.  We adopt here the indices \Hbeta, \MFe, \mgii,
\mgb, \nad\ and \cii\ on which much work has been done in the past so
that the comparison is possible.

\subsection{The Minimum-Distance method}\label{mindist}

As already proposed by TFWG00 the so-called {\it Minimum-Distance
Method} is perhaps best suited to our purposes. Suppose that a galaxy
and/or a star cluster is characterized by a set of observationally
measured indices ${\rm I_{i,obs}}$, from which we want to infer the
age, the metallicity Z, and the degree of \enh\ $\Gamma$. On the
theoretical side, suppose we have the corresponding indices tabulated
as a function of the same quantities in form of discrete grids of
values ${\rm I_{i,th}(t_{j}, Z_{k}, \Gamma_{l})}$ where j, k, and
l vary from 1 to a certain value as appropriate. In the space of the
indices ${\rm I_{i}}$ we define the distance between the observational
set ${\rm I_{i,obs}}$ and a generic point of the correspondent space
${\rm I_{i,th}}$:

\begin{equation}
D = {\rm \left[ \sum _i (I_{i,th}-I_{i,obs})^{2} \right]^{0.5}}
\end{equation}

\noindent
By varying ${\rm I_{i,th}}$ over the whole grid space we find the
particular triplet j, k, l for which D is minimal. This means to
fix the age, metallicity Z, and $\Gamma$.

There are two points of concern with the {\it Minimum-Distance
Method}. Firstly, as pointed out by Buzzoni (2004, private
communication), not all indices are expressed in the same units: some
are in equivalent width others in magnitudes, which poses a physical
inconsistency regarding the distance in such multi-dimensional
space. Furthermore even in the case of homogeneous indices, their
absolute values may greatly differ. In would be better to re-scale all
indices to some suitable units: for instance the variance or the mean
values of the observational sample they are applied to, or the
intrinsic accuracy. In the following we will apply the method as it
is, in order to be able to compare our results with TFWG00. Work is
currently underway to revise the {\it Minimum-Distance procedure}
\citep[][in preparation]{Tantalo04d}. Secondly the method is safe in 
the case of the star clusters because their stellar content is well
mimicked by a SSP. In real galaxies, the situation is more complicate
and risky because even in the case of EGs a mix of stellar populations
with different age and chemical properties is likely to exist.
Therefore the approximation to SSP is no longer valid and SSPs should
be replaced by galactic models incorporating the history of star
formation and chemical enrichment and the indices to be used should
take into account the contribution from all stellar components.
Integrated indices for model galaxies have been calculated by
\citet{Tantalo98b} but never applied to this kind of analysis.
Despite this, since most if not all of the studies in literature are
based on the SSP approximation, we will adopt it also here.

To be applied the {\it Minimum-Distance Method} requires large grids
of SSPs with fine spacing in age, metallicity and $\Gamma$. The SGWC00
grids of SSPs cover a wide age range with narrow spacing, and span a
large range in metallicity, 0.008$\leq$Z$\leq$0.070. Thanks to their
regular behavior over large ranges of age and metallicity additional
SSPs can be added by interpolation. However, the grids are only for
$\Gamma$=0 and $\Gamma$=0.35. The simple extrapolation to higher
values of $\Gamma$ may be risky, because of the non linear response of
the calibration to variations of this important parameter. To cope
with this problem without embarking in the tedious and time consuming
calculations of new grids of stellar models with higher degree of
\enh, the following strategy has been adopted. The detailed stellar
model calculations by SGWC00 have shown that keeping constant all
other physical parameters passing from $\Gamma$=0 to 0.35 or
equivalently re-scaling \FeH\ as appropriate, the stellar models by
themselves do not change in a dramatic fashion but for expected
effects due to the lower \FeH\ and the higher \asfe\ that can be
easily handled. Taking advantage of it, firstly we prepare a new pattern
of chemical abundances having $\Gamma$=0.50 by extrapolating the
variations in individual species passing from $\Gamma$=0 to 0.35
according to the recipe by SGWC00. These have already been shown in
Tables~\ref{enh-deg} and \ref{tab-enh}. Secondly, using the SSPs by
SGWC00 we calculate new grids of indices with the new pattern of
abundances for $\Gamma$=0.50. Finally, we generate the large grid of
SSPs suited to the {\it Minimum-Distance Method}. The grids span the
ranges 0$\leq$T$\leq$20\,Gyr, 0.008$\leq$Z$\leq$0.070 or
--0.40$\leq$[Z/H]$\leq$0.68, and 0$\leq \Gamma \leq$0.50.  The age
spacing is the same as in the SGWC00 database of isochrones. For the
metallicity, by adopting \Zsun=0.019 and \Xsun=0.708, we replace the
abundance by mass Z with [Z/H] so that the results can be immediately
compared to those by TFWG00. The spacing in metallicity is ${\rm
\Delta[Z/H]=0.022}$. Finally, $\Gamma$ is spaced by
$\Delta\Gamma=0.01$.

\subsection{Analysis of the Gonz\'alez sample }\label{result1}

\begin{figure}
\psfig{file=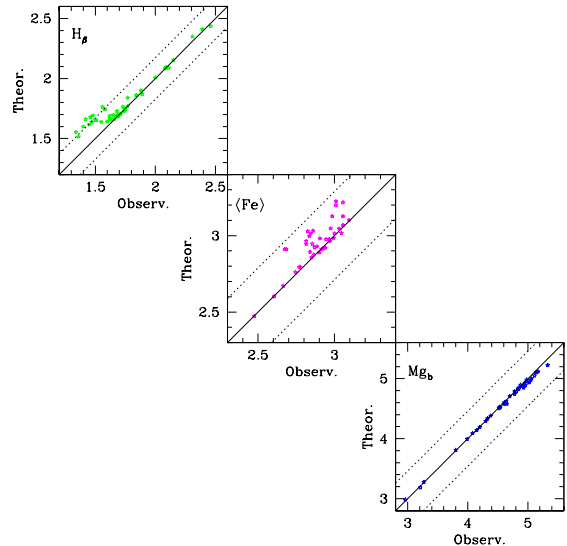,width=8.0truecm}
   \caption{Quality test for the indices \Hbeta, \mgb, \MFe. The
   theoretical value is plotted against the observational one. All
   indices pass the quality test as the theoretical observational
   values coincide within the uncertainty of 10\%.}
\label{quali_gonz}
\end{figure}

Using the above grid of indices, we analyze the same sample of the
\citet{Gonzalez93} catalog of EGs studied by TFWG00 and compare the
results. The indices we have considered are \Hbeta, \mgb\ and \MFe.
The preliminary step is to check if the three indices and the method
pass the quality test, i.e. if the observational input values and the
theoretical ones coincide within an uncertainty of 10\%. The quality
test is shown in the three panels of Fig.~\ref{quali_gonz}.

\begin{figure}
\psfig{file=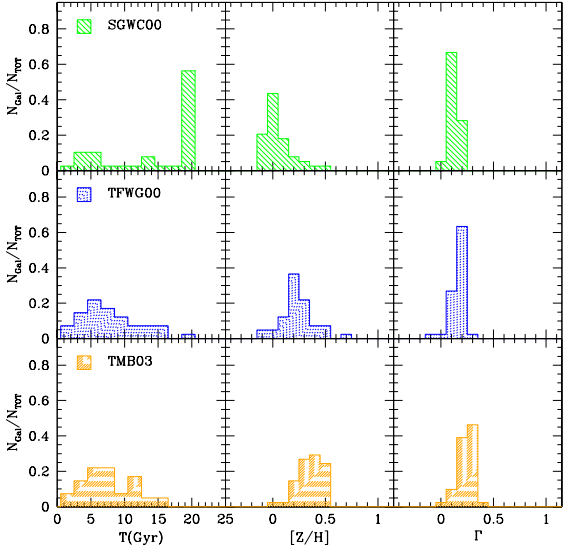,width=8.0truecm}
   \caption[]{Distribution histograms of the age T (in Gyr), metallicity
   ([Z/H]), and enhancement factor $\Gamma$ (from left to right) for
   the galaxies in the Gonz\'alez's sample with ${\rm
   R_{eff}/8}$-aperture. The results are obtained applying the {\it
   Minimum-Distance Method} to the index-triplet \Hbeta, \mgb, \MFe.
   The top panel is for the SGWC00 models, the mid panel is for the
   ${\rm C^{0}O^{+}}$-model of TFWG00. and the bottom
   panel is for the TMB03 models. The last two cases are shown here
   for comparison.}
\label{distgonz}
\end{figure}

\begin{figure}
\psfig{file=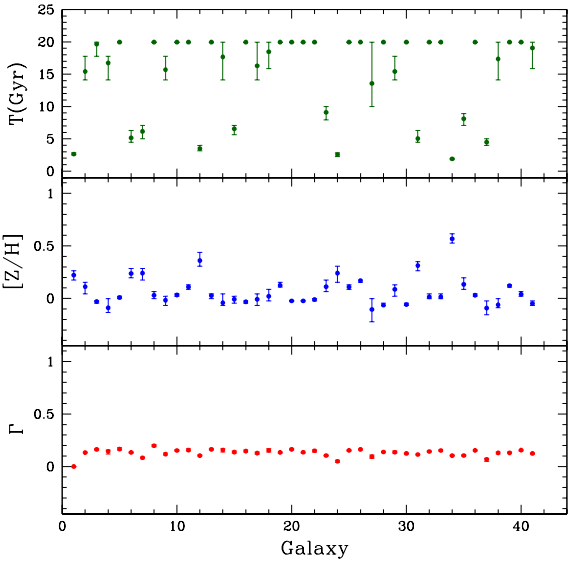,width=8.0truecm}
   \caption{Analysis of the effects due to uncertainty affecting the
   observational indices limited to the results presented in the top
   panel of \protect{Fig.~\ref{distgonz}}. The vertical bars show the
   range spanned by the age, metallicity and $\Gamma$ assigned to each
   galaxy of the \citet{Gonzalez93} list due to the uncertainty
   affecting the observational data. The x-axis is the identification
   number of the galaxies in the catalog. The top panel is the age
   T, the mid panel is the metallicity [Z/H], and the bottom panel
   is the enhancement factor $\Gamma$.}
\label{bar_gonz}
\end{figure}

Fig.~\ref{distgonz} compares the results we obtain here (upper panel)
with those derived by TFWG00 for their ${\rm C^{0}O^{+}}$-model
(central panel), More details on these models are given below. The
various histograms show the fraction of galaxies ${\rm
N_{Gal}/N_{TOT}}$ as a function of the age in Gyr (left panel), the
metallicity [Z/H] (middle panels) and the enhancement factor $\Gamma$
(right panels)\footnote{See Section~9.2 for the correspondence between
our definition of $\Gamma$ and that of [E/Fe] given by TFWG00.}.
Since the observational indices are affected by some uncertainty,
i.e. \Hbeta$\pm \sigma(\Hbeta)$, \mgb$\pm \sigma(\mgb)$ and \MFe$\pm
\sigma(\MFe)$, we derive the uncertainty associated to the theoretical
determinations by searching the grid at [\Hbeta$\pm
\sigma(\Hbeta)$, \mgb, \MFe], [\Hbeta, \mgb$\pm \sigma(\mgb)$, \MFe], and [\Hbeta,
\mgb, \MFe$\pm \sigma(\MFe)$] and taking the maximum deviations,
max($\Delta$T), max(${\rm \Delta [Z/H]}$), and max($\Delta \Gamma$) as
the associated uncertainties. The effect of the observational
uncertainty is measured by the vertical bars on point in the three
panels of Fig.~\ref{bar_gonz}. It is soon evident that the
observational uncertainty on the indices scarcely affects the
determinations of the age, metallicity, and $\Gamma$ but for a few
exceptions. Thanks to it, we will not carry out the analysis of the
uncertainty brought about by the observational errors unless otherwise
specified.

Even considering the uncertainty due to the observational ``errors'',
there are important differences between TFWG00's results and ours, in
particular as far as the age is concerned. While in TFWG00 the
majority of galaxies have ages in the range 1 to 15\,Gyr with a peak
at about 6\,Gyr, in our analysis the majority of galaxies turn out to
be very old. The metallicity distribution peaks at [Z/H]=0
with a long tail up to [Z/H]=0.5. The peak value is about 0.25\,dex
lower than in TFWG00. The distributions of $\Gamma$ are similar. Ours
is only 0.1\,dex lower than in TFWG00.

The major difficulty of the above results is with the age, because too
many galaxies have the formal age of 20\,Gyr, which on one hand is
unacceptably too old on the other hand is the maximum value of the age
grid. This simply means that the solution is poorly determined and all
these cases should be discarded. Arbitrarily dropping all cases for
which the age is older than 15\,Gyr we are left with a handful of
galaxies, whose mean age is about 5\,Gyr.

\citet{ThoMara03} have argued that since SGWC00 $\alpha$-enhanced 
stellar tracks are bluer than the standard ones, they yield higher
\Hbeta\ and weaker indices like \mgii, \mgb, \MFe\ etc. Based on this, 
TMB03 have concluded that the SGWC00 tracks lead to extremely high
ages, without strong impact on metallicity and enhancement factor.
Using the TMB03 grids of theoretical indices applied to the
\citet{Gonzalez93} sample, one gets the results shown in the bottom panel
of Fig.~\ref{distgonz}, which are virtually indistinguishable from
those of TFWG00. Once again the majority of galaxies turn out to be
younger than 10\,Gyr, about 30\% are indeed in the age 10-16\,Gyr;
incidentally 15\,Gyr is the age limit of the TMB03 grids. As far as
the total enhancement and metallicity are concerned, the distribution
of $\Gamma$ peaks in the range $\Gamma$$\simeq$0.2-0.4, whereas the
distribution of metallicities peaks in the range [Fe/H]=0.2 to 0.4. In
any case TMB03 SSPs yield ages that are significantly different from
those of the present study.

The obvious conclusion would be that different stellar models and
different chemical compositions cause the disagreement between TFWG00
and TMB03 results and ours.  We suspect, however, that the reality is
more complex than this simple conclusion. The suspicion arises from
the similarity between TFWG00 and TMB03 despite the large difference
in the input stellar models and SSPs, and the large discrepancy of our
results which provide formal solutions for the age that are at the
limit of the grids. It is therefore mandatory to examine the whole
problem starting from its fundamentals.  The analysis is split in two
parts: firstly (Section~\ref{why}) we will examine in detail the
effects of different stellar inputs (isochrones, SSPs, etc.) and some
important technical details. Secondly in Section~\ref{threecase} we
discuss the effect of different patterns of chemical abundances that
are adopted to build, at given total metallicity Z, the total
enhancement factor $\Gamma$, in other words the effect given the
pattern of ${\rm [X_{el}/Fe]}$.

\section{Why such big differences? The stellar  models and  more}\label{why}

In this section we perform a systematic analysis of the effects on the
absorption line indices diagnostic to assess the galaxy age,
metallicity and degree of enhancement, caused by using SSPs, in which
the various evolutionary phases are taken into account with different
level of accuracy and completeness, by adopting different grids of
theoretical indices to apply the minimum distance method, and finally
by adopting different methods to correct the indices for
$\alpha$-enhancement.

\subsection{The stellar models-isochrones and SSPs}\label{worvspd}

As already mentioned, the indices of TFWG00 are based on the SSPs
calculated by \citet{Worthey94a}. These latter in turn are obtained by
patching together stellar models (isochrones) by \citet{Vandenberg85},
\citet{VandenbergBell85}, \citet{Vanderberg87}, and the Revised Yale
Isochrones \cite[][RYI]{Green87}. The Vandenberg isochrones have been
extended up to the T-RGB phase using the giant branches of RYI with a
number of extrapolations. As far as later stages are concerned, namely
HB and AGB, they have been added by means of the Fuel Consumption
Theorem of \citet{Renbuz86}. See \citet{Worthey92} and
\citet{Worthey94a} for all details. The least we can remark is that
these SSPs are a patchwork of many sources of data, which may not be
fully self-consistent in all details. It is hard, if not impossible,
to trace back all aspects of internal inconsistency (different
physical input in the underlying stellar models, such as opacity,
equation of state, mixing length parameter etc., initial chemical
composition, important details of short lived phases, and finally
numerical accuracy). The SSP models adopted by TMB03 and
\citet{ThoMara03} are those of \citet{Maraston98} which amalgamate
detailed stellar models calculations and the fuel consumption theorem
\citep{Renbuz86} to estimate the energetics of the post main sequence phases. 
Finally, the SSPs of SGWC00 stem from accurate evolutionary stellar
models and isochrones, homogeneous in their input physics and
extending up to the latest visible phases. Their net advantage is (at
least) the internal consistency, which secures that no spurious
effects are added to the problem. 

In addition to this, in the course of our study we will occasionally
use sub-sets of isochrones and SSPs of the Padova library released in
different years in order to compare results based on stellar models
approximately calculated with the same physics (opacity, nuclear
reactions, equation of state, etc). In particular we will use the SSPs
calculated by \citet{Tantalo98a,Tantalo98b}. which are based on
physical input more or less coeval to that adopted for the SSPs of
\citet{Worthey94}. With respect to the classical SSPs by
\citet{Bertelli94}, they differ in the mass-loss rate during the AGB
phase for which the formulation by \citet{Vassiliadis93} was
adopted. Since these SSPs have been amply described for the
first time in the database for galaxy evolution models by
\citet{Leitherer96}, we will refer to them as the 1996 Tantalo's
version of the Padova SSPs (TPD96).

Given these premises, it might be worth of interest (i) to assess the
contribution from stars in different evolutionary phases to the total
value of the SSP indices; (ii) to estimate the uncertainty in the
index values caused by neglecting late evolutionary phase; (iii) to
compare the TFWG00 SSPs with the closest SSPs of the Padova library;
(iv) to examine in some detail the effect of core HeB and later phases
at old ages and/or very high metallicities; (v) to compare results for
the same type of SSPs but different metallicities and enhancement in
$\alpha$-elements; (vi) finally, to mention at least the effect of the
enrichment law $\Delta$Y/$\Delta$Z.
\littleskip 

(i) {\bf Relative contribution by different phases}.
To this aim the SSPs have been split into five evolutionary phases,
i.e. (1) up to the main sequence turn-off (TO), (2) from the TO to the
tip of the RGB (T-RGB), (3) from this to the end of core He-burning
(HeB), (4) from this stage to the end of the TP-AGB, and finally (5)
from this latter down to the formation of Planetary Nebulae and
incipient White Dwarf cooling sequence (P-AGB)\footnote{To be precise
this classification strictly applies to SSPs older than say 0.1\,Gyr,
i.e. those whose TO mass is smaller than about 5\,\Msun.}. We have
considered the SSPs by TPD96 {\it with solar metallicity and no
enhancement in $\alpha$-elements} ($\Gamma$=0). Although these SSPs
are somewhat different from the corresponding ones by SGWC00, they
offer the advantage that the various evolutionary phases have already
been marked by the authors.

\begin{figure}
\psfig{file=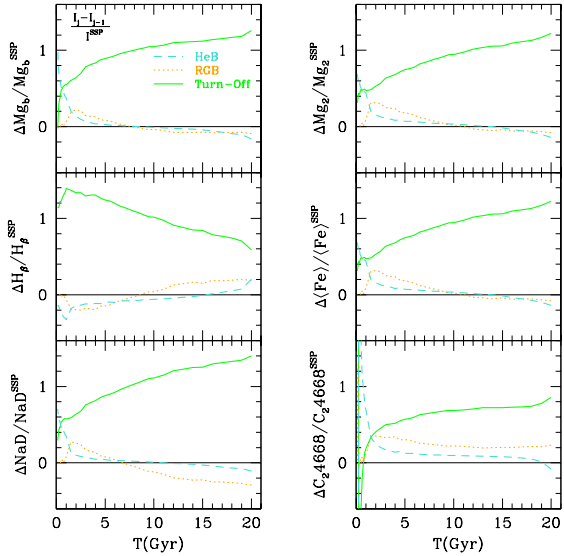,width=8.0truecm}
   \caption{The relative contribution to an index by each phase as
   function of the age (in Gyr). The SSPs are those by TPD96 with
   Z=0.02 and $\Gamma$=0. The {\it solid}, {\it dotted} and {\it
   dashed} lines show the contribution from TO, T-RGB, and HeB phases
   respectively. See the text for more details.}
\label{phase_perc}
\end{figure}

In Fig.~\ref{phase_perc} we show the relative contribution of each 
phase as function of the SSP age. The relative contribution is
defined as follows

\begin{equation}
{\rm 
\left| \frac{\Delta \mathcal{I}_{j}}{\mathcal{I}^{SSP}} \right| = 
\left| \frac{\mathcal{I}_{j}-\mathcal{I}_{j-1}}{\mathcal{I}^{SSP}} \right|
}
\label{single-fase}
\end{equation}

\noindent 
where $\mathcal{I}$ stands for the generic index, $j$ for the phase
running from 1 to 5 ($j-1$=0 is the main sequence),
$\mathcal{I}^{SSP}$ for the total, and finally the phase contribution
$\mathcal{I}_{j}$ is calculated according to eqn.~(\ref{int2_ssp}).

On the average, lumping together the various steps in three major
contributions and neglecting details depending on the particular
index under consideration, we find 
${\rm \Delta \mathcal{I}_{TO}/\Delta \mathcal{I}_{MS}}$=0.1, 
${\rm \Delta \mathcal{I}_{T-RGB}/\Delta \mathcal{I}_{TO}}$=0.3--0.5, 
${\rm \Delta \mathcal{I}_{AGB+P-AGB}/\Delta \mathcal{I}_{T-RGB}}$=0.1. 
\littleskip 

(ii) {\bf Neglecting late evolutionary phases}.
Since not all SSPs in literature contain all evolutionary phases
predicted by the theory of stellar evolution (in some extreme cases
they do not extend beyond the TO or the T-RGB) it is worth looking at
the uncertainty introduced in the indices by neglecting late
evolutionary phases. To this aim we have calculated the indices of
{\it fictitious} SSPs whose last evolutionary phase is the TO, the
T-RGB, the end of core HeB, the end of the TP-AGB, and the P-AGB.
Once again, the SSPs in use are those by TPD96.  Since no other
evolutionary phase is supposed to exist beyond the considered
termination stage, the resulting index is physically consistent even
though it does not correspond to a real value.

\begin{figure}
\psfig{file=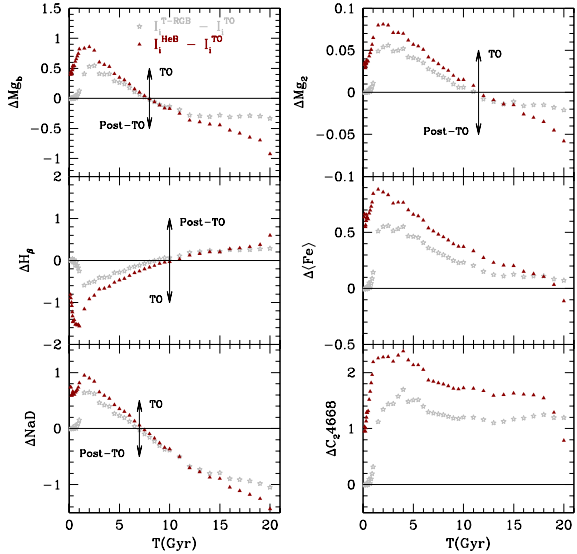,width=8.0truecm}
   \caption{Indices from truncated SSPs, whose last evolutionary stage
   is the TO, the T-RGB, and the end of the HeB, as function of the
   age T in Gyr. The displayed quantities are the differences
   ${\rm \Delta I= I^{T-RGB}-I^{TO}}$ (empty stars) and ${\rm \Delta I=
   I^{HeB}-I^{TO}}$ (filled circles). The SSPs are those by TPD96
   isochrones with Z=0.02 and $\Gamma$=0.}
\label{phase}
\end{figure}

The results of this experiment are shown in Fig.~\ref{phase} which
displays the difference

\begin{equation}
{\rm 
\Delta I= I^{Phase}-I^{TO}
}
\label{ind-diff}
\end{equation}

\noindent
For the sake of simplicity we display only the quantities ${\rm \Delta
I= I^{T-RGB}-I^{TO}}$ and ${\rm \Delta I= I^{HeB}-I^{TO}}$.

It immediately clarifies what follows: (a) The RGB and HeB stages may
significantly affect the total index as already known from the
analysis of the relative contributions; (b) Depending on the age and
index under examination the Post-TO phases may either increase or
decrease the index expected from the sole stars up to the TO. When the
difference is positive, neglecting stages beyond the TO and/or the
T-RGB means that the index in question is underestimated (often by a
significant amount). The opposite when the difference is negative; (c)
In any case all phases beyond core HeB do not significantly contribute
to the final value of an index.

Chief conclusion of the above analysis is that detailed and accurate
calculations of SSPs throughout all evolutionary phases are the main
prerequisite to obtain reliable indices. Passing to the SGWC00 library
of SSPs, we recover most of the trends presented above, even if there
are significant differences in the details.
\littleskip 

(iii){\bf Which of the Padova SSPs get closer to TFWG00? }
Going back to the SSPs adopted by TFWG00, they are not strictly
equivalent to those in usage here first because they are for
solar-scaled compositions and second they stand on stellar models
calculated with somewhat older physical input. To cope with the latter
point of inconsistency, we prefer to compare the SSPs of
\citet{Worthey94} with the TPD96. It is worth recalling that both 
libraries (\citet{Worthey94} and TPD96) even if in a different
fashion, include all stellar evolutionary phases.

\begin{figure}
\psfig{file=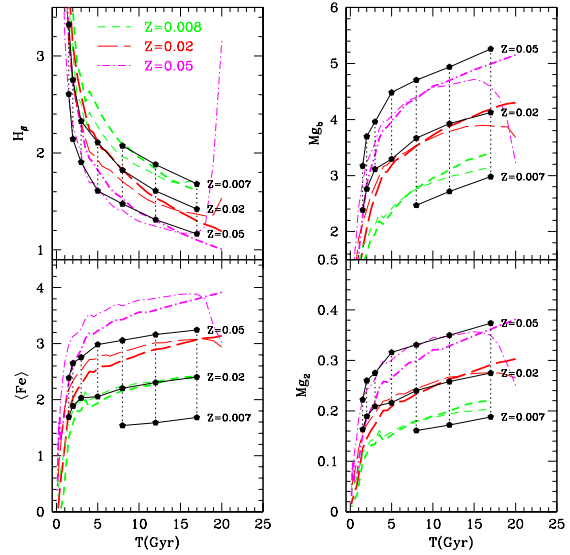,width=8.0truecm}
  \caption{Comparison of four selected indices calculated with
   different SSPs and abundance patterns. All the indices are plotted
   as functions of the age in Gyr and for different metallicities as
   indicated. The thin lines are the indices calculated with the TPD96
   SSPs including all evolutionary phases. The thick lines are the
   same but for SSPs limited to the T-RGB phase. The metallicities are
   Z=0.008, 0.02, and 0.05 ({\it dashed}, {\it long-dashed} and {\it
   dot-dashed} respectively). The full circles joined by thin solid
   and dotted lines are the indices of the\citet{Worthey94a} grid for
   metallicity Z=0.007, 0.02, and 0.05 as indicated.}
\label{worthey}
\end{figure}

In Fig.~\ref{worthey} we compare the indices calculated with the TPD96
SSPs including all evolutionary phases (thin lines) with those
obtained by \citet{Worthey94a} (full dots and lines). The two grids of
models differ in several important details: (a) There is a large
offset in \MFe\ at given metallicity; (b) At old ages the indices
based on TPD96 tend to reverse their trend. The effect is more
pronounced for high metallicities. The reason of this is the anomalous
behavior of the core HeB and later phases at increasing metallicity to
be discussed below. (c) Remarkably, the \citet{Worthey94} grid
supposedly extending up to the latest evolutionary phases is actually
in closer agreement with the TPD96 truncated at the T-RGB. All this
could suggest that part of the disagreement could reside in the way
the HeB stages are included by \citet{Worthey94}. However, the same
argument cannot be invoked for the indices by TMB03, which are based
on stellar models/SSPs that are calculated with accuracy comparable to
that of the TPD96 models.
\littleskip 

(iv) {\bf Effect of core HeB and later phases}.  
Comparing the complete SSPs of TPD96 with those by SGWC00, see
Fig.~\ref{zheh}, we note the important effect on the indices caused by
core HeB stars at varying age and metallicity. In brief, at normal
metallicities, say up to Z$\simeq$0.008, a sort of upper limit for
stars in Globular Clusters, at very old ages (older than about
15\,Gyr) the indices suddenly increase (\Hbeta) or decrease
(e.g. \mgii, \mgb, \MFe). The same occurs for metallicities greater
than Z$\simeq$0.008, but the age at which the trend is reversed gets
lower and lower at increasing metallicity. The reversal of the indices
at old ages and/or high metallicities is fully explained by the
behavior of core HeB stars.  Limiting the discussion to low mass stars
and old ages in turn, the rule is that core HeB (or HB morphology)
takes place closer and closer to the Hayashi line at decreasing age
and/or increasing metal content.  The trend is the result of three
concurring effects: the value of the TO mass (age), the amount of mass
lost at the T-RGB (details of the adopted mass-loss rate are very
important in this context), and the metallicity itself.  Under current
estimates for the efficiency of mass-loss, the HB phase gets redder
and redder at increasing metal content and the AGB phase takes place
along the Hayashi line \citep[see ][ for all details]{Chiosi92}. This
is the typical behavior of stars in Globular Clusters. However, as
pointed out long ago by \citet{Brocato90}, \citet{CasTor91},
\citet{Horch92}, \citet{Dorman93}, \citet{Fagotto94c}, and \citet{Bressan94}, 
if the metallicity happens to be higher than the typical value for the
most metal-rich globular clusters, this simple scheme breaks down. In
brief at high metallicity (close to the solar value), the evolution
does not proceed toward and along the AGB, but toward a slow phase
taking place at high effective temperature without going back toward
the AGB \citep[the so-called AGB-manqu\`e phase, see ][for more
details]{Greggio90}. For solar and twice solar metallicity, the blue
phase begins during the shell He-burning. For 3 times solar
metallicity it begins during the late stages of core He-burning. For
still higher metallicity a large fraction of the core He-burning
lifetime is spent at very high effective temperature and high
luminosity, see \citet{Bressan94} for more details.  All this would
immediately reflect onto the broad-band colors and indices that are
sensitive to the flux emitted in UV-visible part of the spectrum. The
effect is that many indices reverse their trend at increasing age and
increasing metallicity as already shown in Fig.~\ref{worthey} for the
indices \Hbeta, \mgii, \mgb, and \MFe. The SSPs by SGWC00 do not
display the above trend (at least in range of ages and metallicities
they have considered) because of the different physical input of the
stellar models, in particular the mixing length, the opacity, and the
prescription for mass-loss during the RGB and AGB phases (see SGWC00
for details).
\littleskip

\begin{figure}
\psfig{file=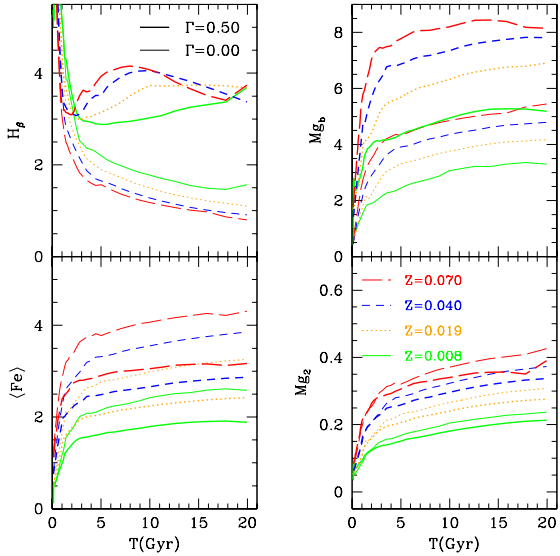,width=8.0truecm}
   \caption{Dependence of the indices \Hbeta, \mgii, \mgb, and \MFe\
   on the age, metallicity and degree of enhancement. The
   metallicities are Z=0.008 (solid lines), 0.019 (dotted-dashed
   lines), 0.04 (dashed lines), and 0.07 (long dashed lines). The thin
   lines are for $\Gamma$=0, whereas the thick lines are for
   $\Gamma$=0.5. The SSPs are those by SGWC00 with the corrections by
   TB95. }
\label{gamma_zeta}
\end{figure}

(v) {\bf Varying metallicity and $\Gamma$}.
Although this topic has already been addressed in
Section~\ref{build_ind} and partially shown in Fig.~\ref{zheh}, it is
worth of interest to stress here once more how the indices, based on
the same type of stellar models, depend on Z and $\Gamma$ considering
the whole range of values spanned by the parameters. The SSPs are
those by SGWC00. The results are shown in Fig.~\ref{gamma_zeta} as
function of the age, metallicity, and $\Gamma$. The thin lines are the
case with $\Gamma$=0, whereas the thick lines are for $\Gamma$=0.5. Of
the four indices on display, \mgii\ has the lowest dependence on
$\Gamma$, the opposite occurs for \mgb. At increasing $\Gamma$, the
index \MFe\ decreases simply due to the fact that higher values of
$\Gamma$ imply lower values of \FeH. The dependence on Z at given
$\Gamma$ is as one would expect from simple considerations. A special
remark is due to \Hbeta, which at increasing $\Gamma$ and Z shows an
unexpected trend. First of all, it significantly increases passing
from $\Gamma$=0 to 0.5. Second, for the case with $\Gamma$=0.5 the
dependence on the metallicity is reversed and even more relevant here
it is no longer monotonic. At ages older than about 3\,Gyr, the more
metal-rich SSPs reach the maximum value in the age range 5 to 10\,Gyr
and then move to lower values. All these trends are in agreement with
the entries of Table~\ref{delta_percent}. Similar results are found
for the SSPs of TPD96.
\littleskip 

(vi) {\bf The enrichment law $\Delta$Y/$\Delta$Z}.
The Padova database of stellar tracks has been calculated assuming a
suitable law of chemical enrichment $\Delta$Y/$\Delta$Z, i.e. the
relation ${\rm Y-Y_{p} = \beta (Z-Z_{p})}$, where ${\rm Y_{p}}$ and
${\rm Z_{p}}$ are the primordial helium and metal abundances,
respectively, and $\beta$ is the enrichment ratio. If ${\rm Z_{p}}$=0
is an obvious choice, excluding the effect of primordial PopIII stars
that in principle could alter both ${\rm Y_{p}}$ and ${\rm Z_{p}}$
\citep{MarChiKud02,Salvaterra03}, the choice of ${\rm Y_{p}}$ and
$\beta$ is more difficult. A recent observational estimate by
\citet{Peimbert02} yields ${\rm Y_{p}}$=0.23 and $\beta$=2.1$\pm$0.5.
\citet{Bertelli94} have adopted ${\rm Y_{p}}$=0.23 and $\beta$=2.5 
\citep{Pagel89}, whereas SGWC00 have chosen ${\rm Y_{p}}$=0.23 and $\beta$=2.25 
as in \citet{Girardi20}. Other databases of stellar tracks and SSPs
have been calculated and/or assembled either explicitly or implicitly
assuming similar enrichment laws, e.g. the SSPs by \citet{Worthey94a}
and TFWG00 in turn have ${\rm Y_{p}}$=0.228 and $\beta$=2.7, whereas
\citet{Maraston98} adopts ${\rm Y_{p}}$=0.23 and $\beta$=2.5 but for 
extremely high metallicities for which adopts the \citet{Salasnich20}
prescription. In many other cases the $\Delta$Y/$\Delta$Z relationship
is simply ignored. Unfortunately no useful set of stellar tracks can
be found in literature, in which a large range of metallicities are
explored at constant Y. Basing on the few and limited set of stellar
tracks to disposal, suffice it to note here that stellar models with
the same Z and higher Y tends to be brighter and bluer than those of
lower Y. This would immediately reflect onto the colors and indices of
the associated SSPs. This is a point to keep in mind when exploring
the effect of Z because depending on the chosen set of models, varying
Z can actually mask also important effects of Y.
\littleskip 

{\bf Final remark}. 
The above discussion clarifies first that not all stellar models are
equivalent -- indeed important differences exist even among libraries
calculated by the same group, see e.g.~TPD96 and SGWC00 (the most
relevant differences here are the opacities and the prescription for
the mass-loss rate at the T-RGB) --, second that details of stellar
models might bear very much on the final correlation between indices
and metallicity and age. This latter point is not trivial considering
that high metallicity stars (well above solar) could exist in the
population mix of EGs, see \citet{Bressan94}. Finally, the effect of
$\Gamma$ is of paramount importance and opposite to what expected from
simple-minded considerations.

\subsection{Comparison between SGWC00 and TMB03}\label{ssp_maraston}

\begin{figure}
\centerline{
\psfig{file=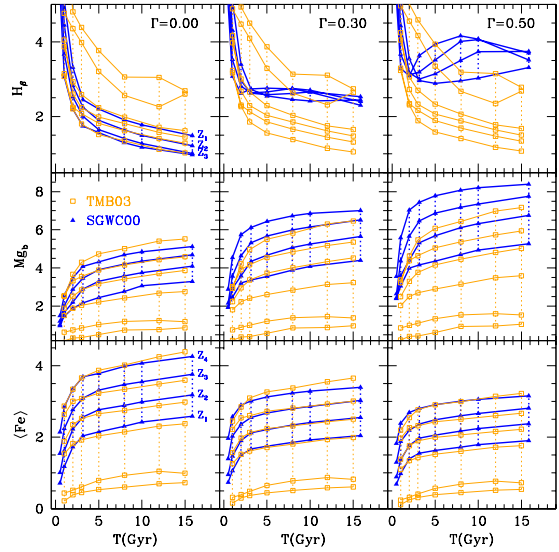,width=8.0truecm}}
  \caption{Comparison between the indices calculated with the SGWC00
  mixture (thick lines) and those by TMB03 with the TFWG00 mixture
  (thin lines). The curves are labelled by the metallicity ${\rm
  Z_{1}}$=0.008, ${\rm Z_{2}}$=0.019, ${\rm Z_{3}}$=0.040, ${\rm
  Z_{4}}$=0.07. The set by TMB03 contains two more metallicities of
  lower value: the top curves in the panels for \Hbeta\ and the bottom
  ones in the panels for the remaining indices. While there is
  coincidence for \mgb\ and \MFe, at increasing $\Gamma$ the values of
  \Hbeta\ disagree.}
\label{ssp_mara}
\end{figure}

In view of the discussion below, it is worth of interest to compare
the indices calculated by TMB03 with ours derived from the SGWC00
stellar models and set of abundances. This is shown in
Fig.~\ref{ssp_mara} for three selected indices, i.e. \Hbeta, \mgb\ and
\MFe, $\Gamma$=0, 0.35 and 0.50, and several values of the metallicity
in common (Z=0.008, 0.019, 0.04, and 0.07). For $\Gamma$=0 the two
sets of models almost exactly coincide, whereas at increasing $\Gamma$
they progressively differ especially as far as \Hbeta\ and \mgb\ are
concerned. For any value of Z, our indices are higher than those of
TMB03. The differences at increasing $\Gamma$ are likely due to either
the adopted enhancement factors for individual elements or the
correcting procedure or both. We will come back to this issue in
Sects.~\ref{threecase} and \ref{Titanium}.

\subsection{Extrapolating the grids of indices}\label{tvspd}

There is another important point to be made that could bear very much
on the final results, i.e. the way grids of stellar
models/SSPs/indices are extrapolated from existing tabulations.

To illustrate the point let us have a close look to the extension made
by TFWG00 of the \citet{Worthey94a} grid of SSPs already shown in the
various panels of Fig.~\ref{worthey}. The maximum age of the SSPs is
17\,Gyr and three values of metallicity are considered, i.e. Z=0.007,
0.02, 0.05. For ages older than about 5\,Gyr the behavior of the
various indices under examination is very smooth, almost linear in the
two parameters. Basing on this, the linear extrapolation of the
indices to older ages, higher metallicities, and higher values of
$\Gamma$ seems to be reasonable. Indeed the grids have been
extrapolated by TFWG00 to ages up to 30\,Gyr and metallicities up to
Z=0.07.

However we have amply discussed that stellar models, SSPs, and indices
in turn are very sensitive to age and metallicity, and that past the
T-RGB some unexpected evolutionary phases may appear largely affecting
the regular trend of the indices in previous stages, at younger ages
and/or lower metallicities. The effect is clear comparing the old
SSPs/indices of TPD96 to those of SGWC00 and the experiments on the
indices that we have performed with the TPD96 SSPs truncated at
different evolutionary phases. Furthermore, we have already noticed
that TFWG00 grids of indices even in the age and metallicity ranges
covered by the original \citet{Worthey94a} models are much akin to the
TPD96 ones truncated at the T-RGB.

Therefore the use of incomplete grids of stellar models and the
straight extrapolation of stellar models/indices to much older ages
and/or much higher metallicities might not be a safe procedure to
adopt.

\subsection{Calculating and correcting SSP indices for $\alpha$-enhancement}\label{corr_indices}

To derive the indices of SSPs and correct them for
$\alpha$-enhancement two different procedures are currently in use:

(i) {\it The star by star case}. This is the method followed by us as
described in Section~\ref{build_ind}: individual stars along an
isochrone of given metallicity and $\Gamma$ are approximated by the
elemental bins of small ${\rm \Delta logL/L_\odot}$,
$\Delta$\Teff. For each bin the indices are derived from the {\it
Fitting Functions} (which in turn depend on \logT, \logG, and [Fe/H]).
The indices of the elemental bins are corrected for enhancement by
means of a suitable technique, and finally the total indices of the
SSP are calculated according to relations~(\ref{int_ssp}).

(ii) {\it The phase by phase case}: TMB03 and TFWG00 split the basic
SSP model in three evolutionary phases: dwarf (D), turn-off (TO), and
giants (G), derive the indices for the each phase summing up the
contribution from the elemental isochrone bin, correct the indices of
each sub-phase by means of the TB95 {\it Response Functions} and
finally get the total indices of the SSP according to
relations~(\ref{int2_ssp}) in Sect.~\ref{def_ind}.

\noindent
To correct the indices for $\alpha$-enhancement by means of the TP95
{\it Response Functions} two possible approximations are in
literature. They have been already been presented in
Sect.~\ref{def_ind}, i.e. relations~(\ref{dind}) and
(\ref{dind_thomas}).

We will see that passing from method (i) plus relation~(\ref{dind}) to
method (ii) plus either relation~(\ref{dind}) as in TFWG00 or
(\ref{dind_thomas}) as in TMB03, the indices will not differ by more
than 15\% provided all other conditions are the same.

\section{Why such big differences? The abundance patterns}\label{threecase}

Another important and obvious source of disagreement are the patterns
of abundances especially when $\alpha$-enhanced mixtures are adopted.
As matter of fact, a star index depends on the gravity, \Teff\ (that
are function of the gross chemical parameters, Y and Z) and detailed
chemical composition at the surface ([Fe/H] at least). This is
particularly relevant when corrections are applied to pass from
solar-scaled to $\alpha$-enhanced mixtures. In other words, an index
is likely to depends more on the detailed pattern of abundances
adopted in the {\it Fitting Functions} and {\it Response Functions}
than on the gross chemical parameters of the underlying stellar
models.

Furthermore, comparing the distributions in age, metallicity and
$\Gamma$ (Fig.~\ref{distgonz}) based on the SGWC00 indices with those
by TFWG00 for their ${\rm C^{0}O^{+}}$ model is not fully correct
because the two patterns of abundances are not the same.

To clarify the subject we perform several experiments combining
different sets of isochrones with different choices for the abundance
patterns, the $\alpha$-enhanced mixtures in particular.

\subsection{{\bf Case A}: The SGWC00 set of abundances}\label{casea}

In this section we remind the reader that a new set of indices
has been calculated adopting the isochrones by SGWC00 and the
total enhancement  $\Gamma$=0.50. Since the abundance pattern of these
models has already been presented in Sect.~\ref{mindist} no more
details are given here. In the following we will refer to the grids of
the SGWC00 indices with $\Gamma$=0, 0.35, and 0.50 as Case {\bf A}
(see the range of this grids in Sect.~\ref{mindist}).

\subsection{{\bf Case B}: The $\rm C^{0}O^{+}$ model of TFWG00}\label{caseb}

Since the TFWG00 indices are calculated from solar-scaled stellar
models-SSPs on the top of which different degrees of enhancement are
added by means of the TB95 calibration (to a first approximation the
effect of enhancement on stellar models is neglected) we have to
recover the same situation. To this aim we adopt the SSPs by TPD96 up
to the P-AGB phases (see section~\ref{worvspd} above), and the pattern
of enhanced abundances as in the model by TFWG00 labelled ${\rm C^{0}
O^{+}}$.

The abundance patterns presented by TFWG00 have three groups of 
elements: 

\begin{description}
\item \underline{Enhanced elements}\footnotemark[6]
\footnotetext[6]{They are scaled up by the same factor. The dependence of the indices 
on Ti is however not taken into account by TFWG00.}: N, Ne, Na, Mg,
Si, S, Ti plus sometimes C and/or O. They are indicated by TFWG00 as
E-elements;

\item \underline{Depressed elements}\footnotemark[7] 
\footnotetext[7]{They are scaled down by the same factor.}(i.e. Fe-peak Group): 
Cr, Mn, Ca, Co, Ni, Cu, Zn, Fe, that we denote as D-elements;

\item \underline{Fixed elements}, which means solar-scaled.
\end{description} 

\noindent 
Furthermore, different values can be assigned to C and O. In their
model ${\rm C^{0} O^{+}}$, C belongs to the group of fixed elements
whereas O to the group of enhanced elements.

To properly compare our results with those by TFWG00, we must
establish the correspondence between our definition of $\Gamma$ with
that adopted by TFWG00. According to their notation, our
eqn.~(\ref{feh1}) can be cast in the following way

\begin{displaymath}
{\rm 
[Fe/H] = [Z/H] -\mathcal{A}[E/Fe]
}
\end{displaymath}

\noindent
or

\begin{displaymath}
{\rm 
\Delta [Fe/H] = - \mathcal{A} \Delta [E/Fe]
}
\end{displaymath}

\noindent
at fixed [Z/H] where ``E'' refers to the mass fraction of elements
that are specifically enhanced and $\mathcal{A}$ is a generic constant
of proportionality. This means that $\mathcal{A}$[E/Fe] corresponds to
$\Gamma$ of eqn.~(\ref{feh1}).

In TFWG00 the enhancement factor is let vary from [E/Fe]=--0.30 to
0.75 which means that the ratio [O/H] increases from --0.021 up to
0.053, whereas [C/H] remains equal to 0. It is worth noticing that the
case with ${\rm [E/Fe]}=-0.30$ and [O/H]=--0.021 considered by TFWG00
actually corresponds to a decrease of the $\alpha$-elements and [O/H]
with respect to the solar pattern.

\begin{table*}
\normalsize
\begin{center}
\caption[]{Element mass fractions in metals for non solar abundance 
ratios according to the model ${\rm C^{0} O^{+}}$ of TFWG00.}
\label{modCO}
\begin{tabular*}{109.5mm}{|c c c c c c c c|}
\hline
\multicolumn{1}{|c}{$\mathcal{A}$} &
\multicolumn{1}{r}{$\Delta$[E/Fe]} &
\multicolumn{1}{r}{$\Delta$[D/H]} &
\multicolumn{1}{r}{$\Delta$[E/H]} &
\multicolumn{1}{c}{${\rm X_{C}/Z}$} &
\multicolumn{1}{c}{${\rm X_{O}/Z}$} &
\multicolumn{1}{c}{${\rm X_{Fe}/Z}$} &
\multicolumn{1}{c|}{${\rm X_{E}/Z}$} \\
\hline
0.9288 & --0.30 &   0.28 & --0.021 & 0.174 & 0.433 & 0.137 & 0.252 \\
0.9288 &   0.00 &   0.00 &   0.000 & 0.174 & 0.482 & 0.072 & 0.265 \\
0.9288 &   0.32 & --0.30 &   0.023 & 0.174 & 0.506 & 0.036 & 0.279 \\
0.9288 &   0.75 & --0.70 &   0.053 & 0.174 & 0.508 & 0.015 & 0.300 \\
\hline
\end{tabular*}
\end{center}
\end{table*}

In Table~\ref{modCO} (the analog of Table 4 in TFWG00) we summarize
the abundance ratios we have adopted to correct the indices according
to their model ${\rm C^{0} O^{+}}$ and our definition of $\Gamma$
given by eqn.~(\ref{feh1}) or (\ref{def2gam}). Column (1) gives the
constant $\mathcal{A}$. Column (2) yields the enhancement factor
([E/Fe] or equivalently $\Gamma/\mathcal{A}$ in our notation). Columns
(3) and (4) list the amount of depression and enhancement for E-
and D-elements, respectively. Finally columns (5) through (8)
give the mass fractions of C, O, Fe and enhanced elements.

The ranges of age, metallicity and enhancement spanned by our new grid
based on the TPD96 SSPs is 0$\leq$T$\leq$20\,Gyr,
0.008$\leq$Z$\leq$0.100 (or --0.42$\leq$[Z/H]$\leq$0.92), and
$-0.28 \leq \Gamma \leq$0.70 (or --0.30$\leq$[E/Fe]$\leq$0.75). The
grid\footnotemark[8] contains four values of metallicity, namely
Z=0.008, 0.02, 0.05 and 0.1 which correspond to [Z/H]=--0.42, 0, 0.47,
and 0.92 according to the TFWG00 notation. For each metallicity we
have corrected the indices by using eqn.~(\ref{dind}) for the
different values of $\Gamma$ given in Table~\ref{modCO}.
Subsequently, we have interpolated among the four grids in steps of
$\Delta \Gamma$=$\mathcal{A}\Delta$[E/Fe]=0.02 or equivalently
$\Delta$[Z/H]=0.03. Finally, the age steps are the same of the
original isochrones. {\it These experiments allow us to test the
effect of a simple enhancement scheme on the standard stellar
models/SSPs}. These models are referred to as Case {\bf B}.

\footnotetext[8]{We adopt \Zsun=0.020 and \Xsun=0.700 to replace Z 
with [Z/H], and the constant $\mathcal{A}$=0.9288 to convert $\Gamma$
into [E/Fe].}

\subsection{{\bf Case C}: The TPD96 isochrones and the SGWC00 mixture}\label{casec}

Finally, we consider the case of indices calculated with the old SSPs
of TPD96 and the mixtures of chemical abundances of SGWC00, i.e.
$\Gamma$=0, 0.35, and 0.50. The grids\footnotemark[9] span the ranges
0$\leq$T$\leq$20\,Gyr, 0.008$\leq$Z$\leq$0.100 (or
$-0.42\leq$[Z/H]$\leq$0.92), and 0.00$\leq \Gamma \leq$0.50. {\it This
case allows us to test the effect of a complex abundance scheme on
standard SSPs}. Models of this type are referred to as Case {\bf C}.
\footnotetext[9]{We adopt \Zsun=0.02 and \Xsun=0.700 to replace Z 
with [Z/H].}

\subsection{Comparing the indices of Cases {\bf A}, {\bf B} and {\bf C}}\label{comp_abc}

\begin{figure}
\psfig{file=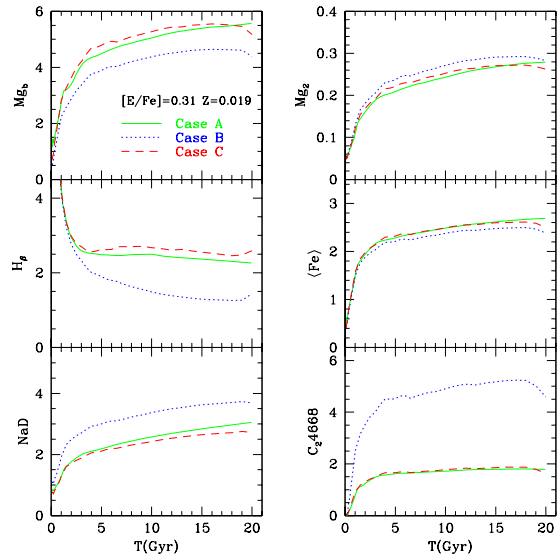,width=8.0truecm}
  \caption{Evolution of six indices (\mgb, \mgii, \Hbeta, \MFe, \nad\
   and \cii) as function of the age according to the recipes of Cases
   {\bf A} (solid lines), {\bf B} (dotted lines), and {\bf C} (long
   dashed lines) and limited to the solar metallicity Z=0.019.}
\label{ind_abc}
\end{figure}

In this section we quickly compare the results for the three cases
above.  In Fig.~\ref{ind_abc} we present the six indices under
consideration as function of the age for the case with solar
composition (Z=0.019) and enhancement [E/Fe]=0.31 (we adopt here the
notation by TFWG00). It is soon evident that Cases {\bf A} and {\bf
C}, whose indices are calculated with same pattern of abundances,
yield nearly the same results despite the fact that they stand on
different SSPs/isochrones. The largest difference is with Case {\bf B}
which has a different set of abundances. This experiment clearly shows
that the abundance ratios are the key parameter. The opposite
conclusion reached by TFWG00 is likely due to an insufficient
exploration of the parameter space since they limit the analysis to
varying only C and O. We show in Section~\ref{Titanium} how the
adoption of a particular set of abundance ratios would reflect onto
the age estimate.
   
\subsection{Ages, metallicities and $\Gamma$s from Cases {\bf A}, {\bf B} and {\bf C}}\label{results}

The results obtained from the {\it Minimum-Distance Method} applied to
the Gonz\'alez's sample and the different sets of theoretical indices
are summarized in Fig.~\ref{alldist}. Uncomfortably, each case yields
different results due to the different assumption for the input SSPs
and pattern of abundances. Looking at the cases in more detail, we
note the following:

\begin{figure}
\psfig{file=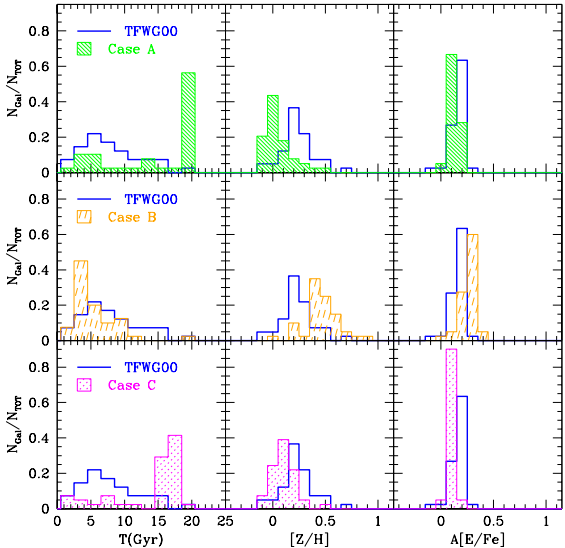,width=8.0truecm}
  \caption{Summary of the results for Cases {\bf A}, {\bf B} and {\bf
  C}. The top three panels show the age (in Gyr), metallicity ([Z/H])
  and enhancement ($\mathcal{A}$[E/Fe]) distributions, from left to
  right for Case {\bf A} models. The thick lines are the TFWG00
  solution, drawn here for comparison. The mid three panels are the
  same but for Case {\bf B} models. Finally the bottom three panels
  are the results for Case {\bf C} models.}
\label{alldist}
\end{figure}

(i) Case {\bf A} -- Ages clusters in two groups, from 1 to 15\,Gyr,
and 19 to 20\,Gyr, where the majority of the population is found.  The
metallicity now goes from ${\rm [Z/H]}=-0.1$ to 0.2 with a tail to higher
values. On the average it is lower than in TFWG00. The enhancement
factor peaks at $\mathcal{A}$[E/Fe]=0.1-0.2 being about 0.1\,dex lower
than in TFWG00.

(ii) Case {\bf B} -- The age distribution is much similar to that of
TFWG00. The only minor difference is the concentration of objects in
the age range 3 to 5 Gyr (about 45\% of the galaxies fall in this age
bin).  The metallicity peaks at about [Z/H]=0.5, and
$\mathcal{A}$[E/Fe]=0.2-0.4. In TFWG00, the ages almost evenly
distribute from 1 to 16\,Gyr with very few objects of 20\,Gyr, the
metallicity is centered at [Z/H]=0.2 with tails on both sides, i.e. a
factor of two lower, the enhancement factor peaks at
$\mathcal{A}$[E/Fe]=0.2, a factor 1.5 lower.

(iii) Case {\bf C} -- The age distribution is much similar to that of
Case {\bf A}, even though the maximum age is now shifted to the range
15-18\,Gyr. The distribution in metallicity is nearly the same as in
TFWG00, only 0.1\,dex lower, whereas the enhancement factor peaks at
$\mathcal{A}$[E/Fe]=0.1 where the majority of galaxies are found with
a short tail extending down to $\mathcal{A}$[E/Fe]=0.

Comparing Case {\bf B} to Case {\bf C} the effect of different
patterns of chemical abundances in the $\alpha$-enhanced mix is
revealed, whereas comparing Case {\bf A} to Case {\bf C}, the effect
of different stellar models/isochrones is highlighted. It is soon
evident that the pattern of abundances in the enhanced mix plays the
dominant role. The results of TFWG00 and TMB03 are indeed recovered
when the abundance pattern is the same, this almost independently of
the SSPs in use.

\begin{figure}
\centerline{
\psfig{file=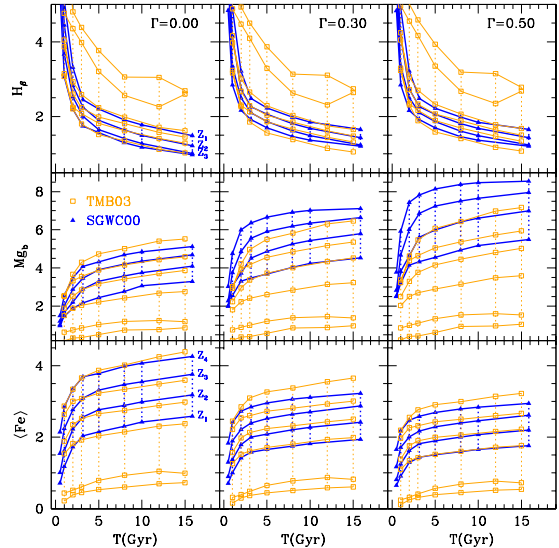,width=8.0truecm}}
  \caption{Comparison of the indices for SSPs with solar metallicity,
   $\Gamma$=0.3 and 0.5, and [Ti/Fe]=0, calculated by
   TMB03 (thin lines) and by us. The curves are labelled by the
   metallicity, i.e. Z$_{1}$=0.008, Z$_{2}$=0.019, Z$_{3}$=0.040,
   Z$_{4}$=0.07. For the metallicities in common, the coincidence is
   remarkably good.}
\label{ssp_noTi}
\end{figure}

\subsection{The nasty Titanium }\label{Titanium}

The lengthy and detailed discussions of the various aspects of the
problem carried out in Sects.~\ref{why} and \ref{threecase} have
highlighted that the abundance pattern for the $\alpha$-enhanced
mixtures is the key parameter. However, pinning down the factor truly
responsible of the disagreement between our study and the previous
ones has so far eluded our efforts.

There is only one aspect of the whole problem, which has not yet been
analyzed, i.e. the detailed comparison element by element for which
variations with respect to the solar mix are adopted.

As already mentioned, at fixed total metallicity SGWC00 adopt a
mixture in which O, Ne, Mg, Si, Ca, Ti, and Ni are enhanced by
different factors with respect to the Sun. All other elements in their
list, specifically C, N, Na, Cr and Fe keep the solar abundance
relative to Fe.

\begin{table*}
\begin{center}
\caption[]{Effects of changing the enhancement factor [Ti/Fe] of Titanium: 
Fractional variations ${\rm \Delta I/I}$ and Indices.}
\label{Ti_resp}
\begin{tabular*}{133mm}{|c l r r r r| c l r r r r|}
\hline
\multicolumn{6}{|c|}{$\Gamma=0.35$} &
\multicolumn{6}{|c|}{$\Gamma=0.50$} \\
\hline
\multicolumn{1}{|c}{[Ti/Fe]} &
\multicolumn{1}{c}{Index} &
\multicolumn{1}{c}{CD} &
\multicolumn{1}{c}{CG} &
\multicolumn{1}{c}{TO} &
\multicolumn{1}{c}{} &
\multicolumn{1}{|c}{[Ti/Fe]} &
\multicolumn{1}{c}{Index} &
\multicolumn{1}{c}{CD} &
\multicolumn{1}{c}{CG} &
\multicolumn{1}{c}{TO} &
\multicolumn{1}{c|}{} \\
\hline 
0.00 & \Hbeta &   0.020 &  -~~~   & --0.003 && 0.00 & \Hbeta & --0.041 &  -~~~   & --0.007 &\\
     & \mgb   &   0.144 &   0.868 &   0.554 &&      & \mgb   &   0.199 &   1.440 &   0.863 &\\
     & Fe52   & --0.166 & --0.230 & --0.201 &&      & Fe52   & --0.232 & --0.314 & --0.294 &\\
     & Fe53   & --0.240 & --0.138 & --0.344 &&      & Fe53   & --0.326 & --0.193 & --0.463 &\\
\hline
0.23 & \Hbeta &   0.802 &  -~~~   &   0.005 && 0.27 & \Hbeta &   0.869 &  -~~~   &   0.003 &\\
     & \mgb   &   0.155 &   0.878 &   0.465 &&      & \mgb   &   0.213 &   1.453 &   0.739 &\\
     & Fe52   & --0.155 & --0.222 & --0.162 &&      & Fe52   & --0.219 & --0.306 & --0.253 &\\
     & Fe53   & --0.221 & --0.128 & --0.316 &&      & Fe53   & --0.307 & --0.182 & --0.436 &\\
\hline
0.43 & \Hbeta &   1.954 &  -~~~   &   0.013 && 0.47 & \Hbeta &   2.065 &  -~~~   &   0.011 &\\
     & \mgb   &   0.165 &   0.885 &   0.392 &&      & \mgb   &   0.223 &   1.464 &   0.653 &\\
     & Fe52   & --0.145 & --0.216 & --0.126 &&      & Fe52   & --0.210 & --0.300 & --0.221 &\\
     & Fe53   & --0.205 & --0.119 & --0.291 &&      & Fe53   & --0.292 & --0.174 & --0.415 &\\
\hline
0.63 & \Hbeta &   3.844 &  -~~~   &   0.021 && 0.67 & \Hbeta &   4.025 &  -~~~   &   0.019 &\\
     & \mgb   &   0.175 &   0.893 &   0.323 &&      & \mgb   &   0.233 &   1.474 &   0.570 &\\
     & Fe52   & --0.135 & --0.210 & --0.089 &&      & Fe52   & --0.201 & --0.295 & --0.188 &\\
     & Fe53   & --0.188 & --0.110 & --0.265 &&      & Fe53   & --0.277 & --0.165 & --0.393 &\\
\hline
     &        &         &         &         && 0.87 & \Hbeta &   7.239 &   0.000 &   0.026 &\\
     &        &         &         &         &&      & \mgb   &   0.244 &   1.485 &   0.492 &\\
     &        &         &         &         &&      & Fe52   & --0.192 & --0.289 & --0.154 &\\
     &        &         &         &         &&      & Fe53   & --0.261 & --0.157 & --0.371 &\\
\hline
\end{tabular*}
\oneskip
\begin{tabular*}{142mm}{|l r c c c c| l r c c c c|}
\hline
\multicolumn{1}{|c}{[Ti/Fe]} &
\multicolumn{1}{c}{Log(T)} &
\multicolumn{1}{c}{\Hbeta} &
\multicolumn{1}{c}{\mgb} &
\multicolumn{1}{c}{$Fe52$} &
\multicolumn{1}{c}{$Fe53$} &
\multicolumn{1}{|c}{[Ti/Fe]} &
\multicolumn{1}{c}{Log(T)} &
\multicolumn{1}{c}{\Hbeta} &
\multicolumn{1}{c}{\mgb} &
\multicolumn{1}{c}{$Fe52$} &
\multicolumn{1}{c|}{$Fe53$} \\
\hline           
0.00&  9.00 & 4.438 & 2.812 & 1.381 & 1.139 & 0.00 &  9.00 & 4.425 & 3.507 & 1.229 & 1.011\\  
    &  9.70 & 1.898 & 5.322 & 2.295 & 2.067 &      &  9.70 & 1.881 & 6.558 & 2.055 & 1.860\\   
    & 10.00 & 1.499 & 5.791 & 2.489 & 2.235 &      & 10.00 & 1.477 & 7.021 & 2.239 & 2.011\\  
    & 10.15 & 1.286 & 6.057 & 2.617 & 2.353 &      & 10.15 & 1.264 & 7.282 & 2.360 & 2.118\\   
\hline
0.23&  9.00 & 4.493 & 2.750 & 1.421 & 1.166 & 0.27 &  9.00 & 4.488 & 3.421 & 1.270 & 1.039\\   
    &  9.70 & 2.065 & 5.243 & 2.346 & 2.110 &      &  9.70 & 2.077 & 6.448 & 2.108 & 1.905\\  
    & 10.00 & 1.753 & 5.723 & 2.539 & 2.283 &      & 10.00 & 1.773 & 6.925 & 2.291 & 2.060\\   
    & 10.15 & 1.552 & 6.003 & 2.665 & 2.403 &      & 10.15 & 1.573 & 7.206 & 2.411 & 2.170\\   
\hline
0.47&  9.00 & 4.554 & 2.700 & 1.456 & 1.191 & 0.47 &  9.00 & 4.551 & 3.362 & 1.302 & 1.061\\  
    &  9.70 & 2.306 & 5.180 & 2.391 & 2.150 &      &  9.70 & 2.326 & 6.373 & 2.148 & 1.939\\   
    & 10.00 & 2.125 & 5.669 & 2.583 & 2.325 &      & 10.00 & 2.159 & 6.860 & 2.331 & 2.097\\   
    & 10.15 & 1.941 & 5.961 & 2.709 & 2.447 &      & 10.15 & 1.976 & 7.155 & 2.450 & 2.209\\ 
\hline
0.63&  9.00 & 4.639 & 2.653 & 1.493 & 1.217 & 0.67 &  9.00 & 4.638 & 3.307 & 1.335 & 1.082\\  
    &  9.50 & 2.709 & 4.582 & 2.263 & 2.027 &      &  9.70 & 2.729 & 6.303 & 2.190 & 1.974\\  
    & 10.00 & 2.731 & 5.618 & 2.629 & 2.369 &      & 10.00 & 2.787 & 6.800 & 2.372 & 2.135\\   
    & 10.15 & 2.576 & 5.922 & 2.753 & 2.492 &      & 10.15 & 2.635 & 7.108 & 2.490 & 2.249\\   
\hline
    &       &       &       &       &       & 0.87 &  9.00 & 4.762 & 3.255 & 1.369 & 1.105\\ 
    &       &       &       &       &       &      &  9.70 & 3.385 & 6.238 & 2.233 & 2.010\\                           
    &       &       &       &       &       &      & 10.00 & 3.813 & 6.745 & 2.414 & 2.174\\
    &       &       &       &       &       &      & 10.15 & 3.712 & 7.065 & 2.531 & 2.289\\   
\hline                                               
\end{tabular*}
\end{center}
\end{table*}

TFWG00 let the abundance of N, Ne, Mg, Na, Si, S to increase, that of
C and O either to increase or to remain constant, that of Fe, Ca, Cr
to decrease, and that of all other elements to remain
constant. However, they have ignored the dependence of the line
strength on Ti as ``TB95 make contradictory statements about its
inclusion in their model atmospheres'' (see the footnote n.~5 in
TFWG00). The mixture adopted by TMB03 is the same of Model 1 by
TFWG00, but they include Ca and Ti in the group of enhanced elements.

Looking at the {\it Response Functions} of TB95, those of Mg, and Ti
are the dominants ones in particular for the Cool Dwarfs. For an
enhancement factor 0.3, the TFWG00 and TMB03 mixtures imply
[Mg/H]=0.023 and [Ti/H]=0.023. Both are significantly smaller than our
values listed in Table~\ref{tab-enh}, the one for Ti in particular
which is a factor of ten smaller. We expect therefore that differences
in the degree of enhancement adopted for these elements should bear
very much on the final results. To check this point we have
systematically changed the quantities ${\rm [X_{el}/Fe]}$ and derived
the relative variations ${\rm \Delta I/I}$ for the Cool-Dwarf,
Cool-Giant, and Turn-Off stars of TB95. As expected Mg and Ti have a
large impact, whereas all remaining elements do not play a significant
role. Since Mg is always enhanced by a comparable factor in all
studies under examination, whereas Ti is strongly enhanced only in
SGWC00, we suspect that the ultimate cause of the disagreement between
TFWG00, TMB03 and the present study is the degree of enhancement
assumed for this element. Table~\ref{Ti_resp} summarizes the results
of the analysis. The overabundance factor [Ti/Fe] is increased in
steps of 0.2\,dex from [Ti/Fe]=0 to the value adopted by SGWC00 for
$\Gamma$=0.35, i.e. [Ti/Fe]=0.63, and to the value [Ti/Fe]=0.87
derived by us for $\Gamma$=0.50. Limited to the case of SSPs with
Z=0.019 (similar results are found for other metallicities), the
fractional variations ${\rm \Delta I/I}$ and indices are
calculated. While the variations for indices like \mgb, Fe5270, Fe5335
are nearly independent of [Ti/Fe], $\Delta\Hbeta/\Hbeta$ greatly
increases passing from [Ti/Fe]=0 to higher and higher values. This
immediately reflects onto the value of \Hbeta\ which follows the same
trend in particular for ages older than 1\,Gyr. {\it Is the
overabundance of Ti the sole cause of the disagreement between TFWG00,
TMB03 and SGWC00?}

To answer this question, we have calculated a new grid of SSPs for
which we assume that [Ti/Fe]=0 in the SGWC00 mix and also in our case
with $\Gamma$=0.50. We compare the time variation of a few selected
indices with those by TMB03. The results are shown in
Fig.~\ref{ssp_noTi}. The agreement is now remarkably good. There is a
marginal difference at old ages amounting to the maximum value of
15\%, which, as already mentioned, is caused by the correcting
procedure, i.e. the star by star versus the phase by phase technique,
and the use of relation~(\ref{dind}) instead of relation
(\ref{dind_thomas}).

\begin{figure}
\centerline{
\psfig{file=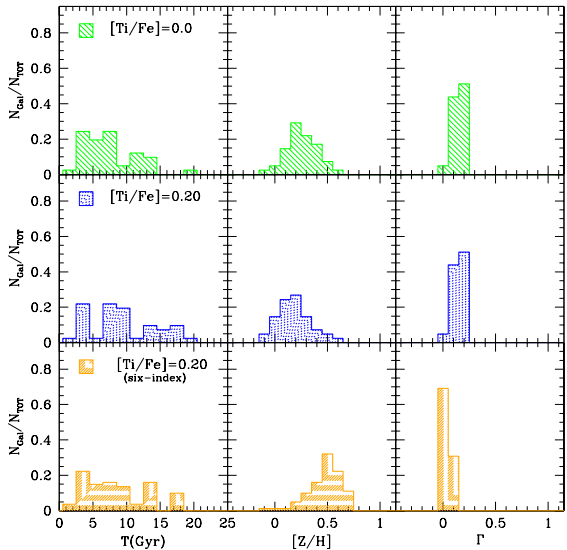,width=8.0truecm}}
  \caption{{\bf Top Panels}: Solutions for models with [Ti/Fe]=0. The
  list of enhanced elements is the same as in SGWC00 and Case {\bf A}
  models, but for Ti, for which the solar value is adopted. But for
  one exception all galaxies are younger than 15\,Gyr. The results
  refer to the \citet{Gonzalez93} galaxies. The solution is
  essentially the same as in TFWG00 and TMB03.  {\bf Middle Panels}:
  Solutions for Case {\bf D} models with [Ti/Fe]=0.20. Even with the
  small enhancement factor of Globular Clusters, the solution tend to
  go back to the starting case presented in Fig.~\ref{distgonz}. The
  distributions refer to the same galaxies as in the Top Panels.  {\bf
  Bottom Panels}: Solutions from Case {\bf D} models with [Ti/Fe]=0.20
  but using six indices instead of three. Now the distributions are
  for the \citet{Trager97} catalog.}
\label{trivial}
\end{figure}

It easy to foresee that the age, metallicity, and enhancement
histograms will not differ too much from those already presented by
TFWG00 and TMB03. Since the exercise is trivial we will not go into
any detail. Suffice to show in the top panels of Fig.~\ref{trivial}
the results we get for the \citet{Gonzalez93} sample. No further
comments are required here.

The reason for the very old ages we have found for most
galaxies using the SGWC00 indices are the much higher values and the
non-monotonic age-dependence of \Hbeta\ (which usually decreases at
increasing age) for $\alpha$-enhanced indices, which in turn are due
to the adopted high values of [Ti/Fe]. The other indices in use
are scarcely affected by [Ti/Fe] and always keep a monotonic age
dependence. Since most observational \Hbeta\ fall in the range 1 to
2, rarely up to 2.5, in a theoretical parameter space favoring high
{\Hbeta}s, while leaving unchanged the remaining indices, the
observational triplet of indices is rendered by pushing the solution
(in particular for the age) toward the old age edge of the grid
looking for low {\Hbeta}s. Indeed the most common solution is ``very
old ages, solar metallicity, and weak enhancement'' (see
Fig.~\ref{distgonz}).

\subsection{What can we do?}

Given the extreme sensitivity of the results to the enhancement
factors, the one for Ti in particular, it is worth comparing the
choice made by SGWC00 with the most recent determinations obtained by
\citet{Gratton03} for field stars with accurate parallaxes.
The mean enhancement for all elements lumped together estimated by
\citet{Gratton03} is ${\rm [\alpha/Fe]}\simeq$0.30 (i.e. $\Gamma$=0.30 
in our notation), which roughly corresponds to the case $\Gamma$=0.35
in SGWC00. The enhancement factors for the two sources of data are
listed in Table~\ref{overgratton}. The ratios ${\rm [X_{el}/Fe]}$ for
O, Mg, Si, Ni, Na, Cr, and Fe are almost coincident; that of Ca
differs by 0.24\,dex, nothing can be said for Ne and S because they
are not included in the \citet{Gratton03} list; finally there is large
difference for Ti, which is [Ti/Fe]=0.20 in \citet{Gratton03} and 0.63
in SGWC00. Although there is no compelling evidence that the data for
Galactic field stars hold good also for elliptical galaxies, for which
the enhancement factors could be different (likely higher), we may
take the estimate of \citet{Gratton03} for [Ti/Fe] as an indicative
value for $\Gamma$=0.3 and use it to predict the abundances for
$\Gamma$=0.5. Therefore, we replace [Ti/Fe] of SGWC00 with this value,
keep unchanged the enhancement factor for the other species in the
list, rescale the abundances as appropriate, and calculate a new grid
of indices therein after referred to as Case {\bf D}. Repeating the
analysis for the age, metallicity and enhancement, we get the results
presented in the middle panels of Fig.~\ref{trivial}. Although the
situation is not the same as in Fig.~\ref{distgonz}, it is also
somewhat different from the one presented in the top panel of
Fig.~\ref{trivial} because a significant fraction of galaxies with
ages from 15 to 18\,Gyr are found. This once more confirms the
sensitivity of the results to the adopted [Ti/Fe]. Lower values for
this parameter are likely more appropriate. In spite of this, Case
{\bf D} indices will used throughout the second part of this study.

\begin{table*}
\normalsize
\begin{center}
\caption[]{Overabundances in SGWC00 and \citet{Gratton03}.}
\label{overgratton}
\begin{tabular*}{166.5mm}{|c| c| c| c| c| c| c| c| c| r| c| c| c| r| c|}
\hline
\multicolumn{1}{|c}{${\rm [X_{el}/Fe]}$} &
\multicolumn{1}{|c}{$\alpha$} &
\multicolumn{1}{|c}{O} &
\multicolumn{1}{|c}{Ne} & 
\multicolumn{1}{|c}{Mg} &
\multicolumn{1}{|c}{Si} & 
\multicolumn{1}{|c}{S} &
\multicolumn{1}{|c}{Ca} &
\multicolumn{1}{|c}{Ti} &
\multicolumn{1}{|c}{Ni} &
\multicolumn{1}{|c}{C} &
\multicolumn{1}{|c}{N} &
\multicolumn{1}{|c}{Na} &
\multicolumn{1}{|c}{Cr} &
\multicolumn{1}{|c|}{Fe} \\
\hline
SGWC00         & 0.35 & 0.50 & 0.29 & 0.40 & 0.30 & 0.33 & 0.50 & 0.63 &   0.02 & 0.00 & 0.00 & 0.00 &   0.00 & 0.00 \\
Gratton~et~al. & 0.30 & 0.51 &  -   & 0.39 & 0.25 &  -   & 0.26 & 0.20 & --0.03 & 0.00 & 0.00 & 0.02 & --0.02 & 0.00 \\
\hline
\end{tabular*}
\end{center}
\end{table*}

\section{Is the solution unique?}\label{unique}

Another important question to be addressed is whether different
indices yield the same answer as far as age, metallicity and enhancement
factor are concerned. In so far we have analyzed the triplet \Hbeta,
\MFe\ and \mgb\ because tighten to the Gonz\'alez sample. However,
several equally representative catalogs of galactic indices are
available in literature, for instance the Trager ``{\it IDS
Pristine}'' sample \citep{Trager97}, which allow us to derive the
above parameters first for a larger number of galaxies and even more
relevant here for different groups of indices. In the following we
will consider six different indices (\mgb, \mgii, \Hbeta, \MFe, \nad\
and \cii) and all possible combinations in groups of three. The
inclusion of \nad\ and \cii\ deserves some cautionary
remarks. According to \citet{Thomas03} both indices are not well
calibrated. In addition to this, \cii\ is very sensitive to the C
abundance (TB95) and \nad\ is contaminated by interstellar absorption
\citep{Maraston03}. The only advantage with these indices is their 
sensitivity to Z and $\Gamma$ (see the entries of Table~\ref{delta_percent}).
Keeping these caveats in mind, we decided to make use of these indices,
\cii\ in particular, because looking at the quality ranking we are going 
to present, the situation is not as bad as it may seem. The
observational values are indeed well reproduced by their theoretical
counterparts.

An index is considered to be eligible for this kind of analysis if its
observational and theoretical value coincide within an uncertainty of
10\%. In Table~\ref{indtriple_D} we summarize the results of the
ranking analysis for the Case {\bf D} indices, where Y means that a
good correlation between the observational and the theoretically
recovered value is found, whereas N is the opposite. Three triplets
fully pass the ranking text, i.e. \Hbeta-\mgb-\mgii, \Hbeta-\mgb-\cii,
and \mgb-\nad-\cii

\begin{figure}
\centerline{
\psfig{file=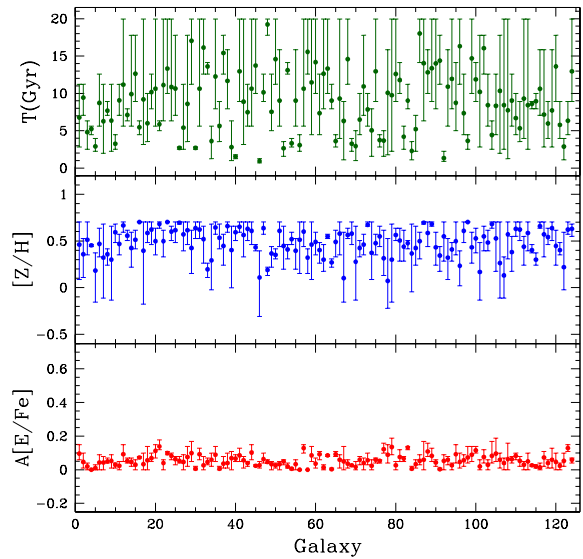,width=8.0truecm}}
  \caption{Age, metallicity, and enhancement factor obtained from the
  Case {\bf D} models but using different index-triplets,
  i.e. \Hbeta-\mgb-\mgii, \Hbeta-\mgb-\cii, and \mgb-\nad-\cii. The
  full circles show the mean value of age, metallicity and enhancement
  factor estimated for each galaxy of the Trager ``{\it IDS
  Pristine}'' sample whereas the vertical bars indicate the maximum
  and minimum values. Different triplets yield different results for
  most galaxies.}
\label{fig_caseD}
\end{figure}

The analysis is made for all the galaxies of the \citet{Trager97}
catalog using the {\it Minimum-Distance Method}. The results for the
distribution of age, metallicity and enhancement factor are displayed
in the various panels of Fig.~\ref{fig_caseD}. Each galaxy is
identified by its list number in the \citet{Trager97} catalog. In each
panel we show the range (the vertical bar) spanned by the
determinations and their mean value (the full circle) obtained from
the different triplets of indices. The top, mid and bottom panels
show the age, the metallicity, and the enhancement factor,
respectively. It is soon evident that a large spread exists for the
same parameter derived from different triplets of indices.

Is the dispersion real? In the sense that each combination of indices
is more sensitive to some specific property of the underlying stellar
mix and therefore it traces the stellar component most contributing to
the indices in question. Or despite the merit parameter not all
triplets are actually good indicators?

\begin{table}
\normalsize
\begin{center}
\caption[]{Ranking analysis of the index-triplets for Case {\bf D} models.}
\label{indtriple_D}
\footnotesize
\begin{tabular*}{81mm}{|l |c |l |c|}
\hline
\multicolumn{1}{|c|}{Index-Triplet} &
\multicolumn{1}{c|}{Case {\bf D}} &
\multicolumn{1}{c|}{Index-Triplet} &
\multicolumn{1}{c|}{Case {\bf D}} \\
\hline
& & & \\
${\rm H_{\beta} ~\langle Fe \rangle ~Mg_{b}}$    & N Y Y & ${\rm \langle Fe \rangle ~Mg_{b} ~Mg_{2}}$       & Y Y N \\
& & & \\
${\rm H_{\beta} ~\langle Fe \rangle ~Mg_{2}}$    & N Y N & ${\rm \langle Fe \rangle ~Mg_{b} ~NaD}$          & N Y Y \\
& & & \\
${\rm H_{\beta} ~\langle Fe \rangle ~NaD}$       & N N N & ${\rm \langle Fe \rangle ~Mg_{b} ~C_{2}4668}$    & N Y Y \\
& & & \\
${\rm H_{\beta} ~\langle Fe \rangle ~C_{2}4668}$ & N N Y & ${\rm \langle Fe \rangle ~Mg_{2} ~NaD}$          & N N Y \\
& & & \\
${\rm H_{\beta} ~Mg_{b} ~Mg_{2}}$                & Y Y Y & ${\rm \langle Fe \rangle ~Mg_{2} ~C_{2}4668}$    & N N Y \\
& & & \\
${\rm H_{\beta} ~Mg_{b} ~NaD}$                   & N Y Y & ${\rm \langle Fe \rangle ~NaD ~C_{2}4668}$       & N Y Y \\
& & & \\
${\rm H_{\beta} ~Mg_{b} ~C_{2}4668}$             & Y Y Y & ${\rm Mg_{b} ~Mg_{2} ~NaD}$                      & Y N Y \\
& & & \\
${\rm H_{\beta} ~Mg_{2} ~NaD}$                   & N N Y & ${\rm Mg_{b} ~Mg_{2} ~C_{2}4668}$                & Y N Y \\
& & & \\
${\rm H_{\beta} ~Mg_{2} ~C_{2}4668}$             & Y N Y & ${\rm Mg_{b} ~NaD ~C_{2}4668}$                   & Y Y Y \\
& & & \\
${\rm H_{\beta} ~NaD ~C_{2}4668}$                & N N Y & ${\rm Mg_{2} ~NaD ~C_{2}4668}$                   & N Y Y \\
& & & \\
\hline
\end{tabular*}
\end{center}
\end{table}

We have performed many numerical experiments using fictitious SSPs
whose terminal stage is the TO, the T-RGB, and the P-AGB in order to
understand if the above dispersion is caused by a different response
of triplets of indices to some specific evolutionary phase or group of
stars. They are not shown here for the sake of brevity. Even if some
indices seem to be weakly sensitive to some particular evolutionary
phases, the large differences in age, metallicity and enhancement
factor that are always found at varying the triplet of indices in use
simply reflect that the solution is not firmly established.  This is
point of embarrassment because there are no strong arguments to prefer
one triplet with respect to others. The
dispersion we find means that the solution is not firmly constrained.

To cope with this point of difficulty a last attempt is made using all
the six indices at once applied to Case {\bf D} models. The results
are shown in the bottom panels panel of Fig.~\ref{trivial}. Compared
to the results for the same models obtained from the three indices
\Hbeta, \mgb, and \MFe\ (middle panels), the number of
galaxies of very old age has decreased in favor of those with young
age (more 90\% of the total); the metallicity is on the average
0.3\,dex higher, and $\Gamma$ goes from zero to about 0.1.

\begin{figure*}
\centerline{
\psfig{file=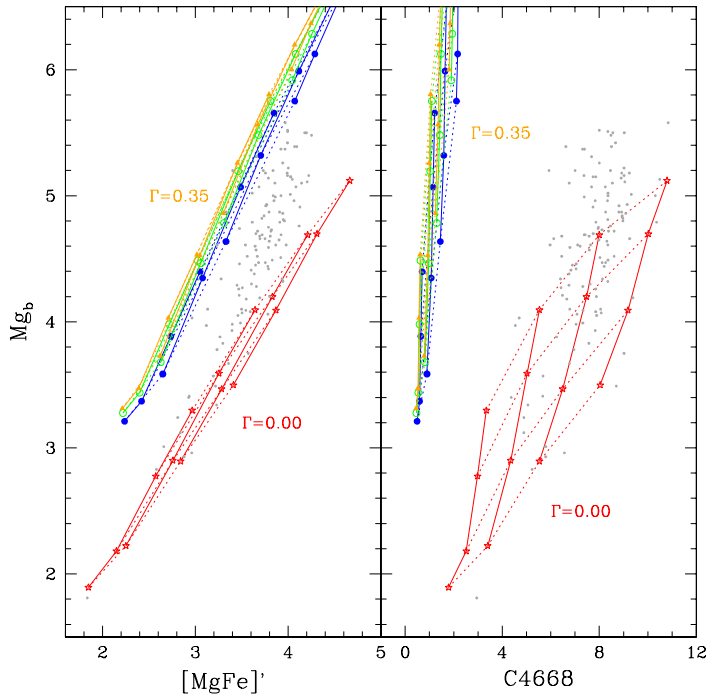,width=10.0truecm}}
  \caption{{\bf Left Panel:} The \MgFeb\ vs. \mgb\ plane for
  $\Gamma$=0 and $\Gamma$=0.35 as indicated. In each group four
  metallicities are shown, i.e. Z=0.008, 0.019, 0.04, 0.07 (from left
  to right). For each $\Gamma$ and metallicity, we show the results
  for [Ti/Fe]=0 (filled triangles), [Ti/Fe]=0.20 (empty circles), and
  [Ti/Fe]=0.63 (filled circles). The dashed lines are the loci of
  equal age. The theoretical indices on display are those of Case {\bf
  D}. The small dots are the data for the Trager ``{\it IDS
  Pristine}'' sample. {\bf Right Panel:} The same but for the \mgb\
  vs. \cii\ plane.}
\label{grid_1}
\end{figure*}

\section{A few remarks on the two indices diagnostic}\label{twoind}

The two-indices planes are customarily used to interpret the
observational data of galaxies and star clusters. Popular planes are
the \Hbeta\ or \hgf\ vs. \MFe\ or \MgFe\ or \mgb\ to derive the age
and mean metallicity \citep{Bressan96,Gonzalez93,Trager20a}, and the
\MFe\ vs. \mgii\ to asses the degree of enhancement
\citep{Worthey92,Gonzalez93,Weiss95}. The two-indices diagnostics 
suffers the same uncertainty encountered with the {\it
Minimum-Distance Method}, however at a lower level of complexity
because some of the parameters are hidden.

In the following we will examine three typical planes which associate
indices that are more sensitive (or unsensitive) to some specific
parameter. For instance \MgFe\ and \MgFeb\ do not depend on $\Gamma$,
the index \cii\ is very sensitive to Z, partially to $\Gamma$ and age,
and so forth.

\begin{figure*}
\centerline{
\psfig{file=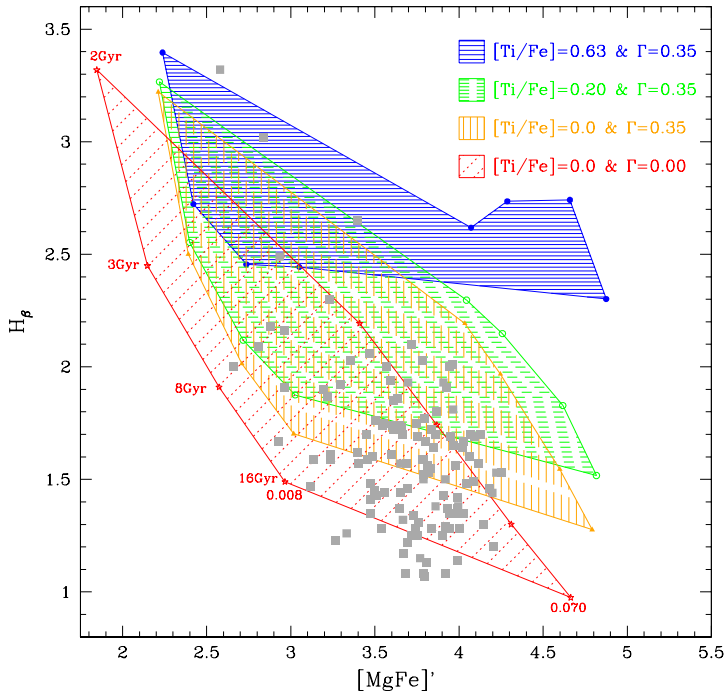,width=10.0truecm}}
  \caption{The \Hbeta\ vs. \MgFeb\ plane for different combinations of
  $\Gamma$=0, metallicity, and [Ti/Fe] as indicated. Each hatched area
  is enclosed between the two SSPs with the lowest and highest
  metallicity in our sample, i.e. Z=0.008 (left) and 0.07
  (right). Along each SSP four values of the age are marked so that
  the lines of constant age can be drawn (they are not shown in here
  for the sake of clarity but for the cases of 2 and 16\,Gyr). The
  different hatched areas correspond to different combinations of
  $\Gamma$ and [Ti/Fe]. The theoretical indices on display are those
  of Case {\bf D}. The large filled squares are the data from the
  Trager ``{\it IDS Pristine}'' sample.}
\label{grid_2}
\end{figure*}

In Fig.~\ref{grid_1} we show the planes \mgb\ vs. \MgFeb\ (left panel)
and \mgb\ vs. \cii\ (right panel). In each panel are displayed the
indices for two values of $\Gamma$ (0 and 0.35), three values of
[Ti/Fe], i.e. 0, 0.20 and 0.63 (filled triangles, empty and filled
circles, respectively), and four values of the age, i.e. 16, 8, 3 and
2\,Gyr. In Fig.~\ref{grid_2} we do the same but for the plane \Hbeta\
vs. \MgFeb.

(i) {\it The \mgb\ vs. \MgFeb\ plane}: this relationship strongly
depends on $\Gamma$ via the index \mgb\ whereas there is little
resolving power at varying Z, [Ti/Fe] (see the entries of
Table~\ref{Ti_resp}) and age. The total enhancement factor of the
Trager galaxies goes from $\Gamma$=0 to 0.35 with mean value
$\Gamma$=0.10. Nothing can be said for the remaining parameters.

(ii) {\it The \mgb\ vs. \cii\ plane}: this relationship strongly
depends on the metallicity and age for $\Gamma$=0 (solar) whereas it
gets worst (little resolving power) for $\Gamma$=0.35. The data
cluster in the metallicity range 0.02$\leq$Z$\leq$0.04 (with a few
exceptions on both sides), are likely compatible with $\langle \Gamma
\rangle$=0.10 (values of $\Gamma$ as high as 0.35 are likely to be
excluded), and seem to span a wide range of ages. Incidentally, the
low values of $\Gamma$ agree with the estimates from the {\it
Minimum-Distance Method} whereas the metallicity is about 0.2\,dex
lower (see Figs.~\ref{trivial} and \ref{fig_caseD}).

(iii) {\it The \Hbeta\ vs. \MgFeb\ plane}: this is the most
interesting and complex plane to look at. In this plane we have
plotted the relationships at varying metallicity, $\Gamma$, [Ti/Fe],
and age. Each hatched area is enclosed between the two SSPs with the
lowest and highest metallicity in our sample, i.e. Z=0.008 (left) and
0.07 (right). Along each SSP four values of the age are marked, so
that the lines of constant age can be drawn (they are not shown in
here for the sake of clarity). The different hatched areas correspond
to different combinations of $\Gamma$ and [Ti/Fe]. Starting from the
solar case ($\Gamma$=0 and [Ti/Fe]=0) we plot the cases $\Gamma$=0.35
and [Ti/Fe]=0, 0.20, and $\Gamma$=0.35 and 0.63. The case
$\Gamma$=0.50 is of little interest here. Finally, the filled squares
are the data.  We emphasize here that other elements like Ti could
play the same role: the only requirement is that they have positive
fractional variations ${\rm (\Delta I/I)}$ to increasing $\Gamma$ and
${\rm [X_{el}/Fe]}$.  Assigning age, metallicity and degree of
enhancement to a galaxy is a cumbersome affair, because all the three
parameters together with possible differences in the abundance ratios
${\rm [X_{el}/Fe]}$ of some elements concur to scatter the data across
this plane. In other words,  a galaxy of given \Hbeta\ and \MgFeb\
could have a certain age and a certain metallicity in absence of
enhancement, be older and less metal-rich in presence of enhancement,
however in this latter case the true age and metallicity depending
also on the abundance ratios ${\rm [X_{el}/Fe]}$ of some elements (the
case being represented here by the effect of [Ti/Fe]). In addition,
the large spread along the \Hbeta\ axis, which is customarily
interpreted as an age spread, could be due to a spread in chemical
abundances and abundance ratios. Not necessarily, a galaxy with strong
\Hbeta\ is a young object, The situation becomes very embarrassing and
interesting at the same time because another dimension is added to the
problem: the spread in \Hbeta\ could be the signature of different
star formation histories and chemical enrichment in turn from galaxy
to galaxy. Owing to its implications the issue has been the subject of
a companion paper by \citep[][]{Tantalo04b}.

\section{Summary and concluding remarks}\label{concl}

In this study we have investigated the ability of absorption line
indices to assess the metallicity Z, the iron content \FeH\ and
$\Gamma$ in turn, and the age of elliptical galaxies. The analysis has
been developed through several steps:

(i) Firstly, we have generated grids of indices based on the SSPs by
SGWC00 calculated with chemical composition and opacities enhanced in
$\alpha$-elements, and transformations in which the same chemical
mixtures have been adopted. The SSPs and associated indices are for
three degrees of total enhancement, i.e. $\Gamma$=0, 0.35, and 0.50.

(ii) Secondly, we have assessed the response of the indices to
variations of age, metallicity, and $\Gamma$ and found that most of
the indices have similar response to three parameters but for a few of
them which are definitely more sensitive to metallicity and
enhancement. It is worth calling attention that most indices are
evenly and weakly depending on the age over large ranges. This is the
most crucial aspect of the problem.

(iii) Applying the new grids of SSPs/indices to the popular sample of
EGs by \citet{Gonzalez93} we have found distributions of age,
metallicity and enhancement factors that were too different with
respect to previous studies (e.g. TFWG00 and TMB03) of the same data
to be simply explained in a simple fashion. This spurred a systematic
analysis of the whole problem, in which we have examined the effect of
the stellar models and SSPs in use, of the contribution to the
integrated indices of SSPs by stars in different evolutionary phases
(TO, T-RGB, HeB, TP-AGB, and later), of the evolutionary behavior of
stars with unusual metallicity, the enrichment law
$\Delta$Y/$\Delta$Z, of technical details such as the extrapolation of
existing grids of theoretical indices to ages, metallicities,
enhancement factors not covered by the original models. Even if each
of them bears on the final results. Lastly, we have examined the
effect of the particular set of abundance ratios ${\rm [X_{el}/Fe]}$
that is adopted to reach a certain degree of total enhancement
($\Gamma$) at fixed metallicity Z.

(iv) This poses another crucial question because the results much
depend on the specific way in which the abundance pattern is built up
at given $\Gamma$ and Z. We have clarified that main reason for the
difference between the results obtained with the SGWC00 models and
those by TFWG00 and TMB03 is the different assumption made for the
ratio [Ti/Fe]. The difference originates from the high {\it Response
Function} for this element. We suspect that other elements with high
{\it Response Function} would yield similar effects.  This finding
adds other dimensions to the problem because the assignment for age,
metallicity and $\Gamma$ much depend on the ratios ${\rm [X_{el}/Fe]}$
at least from important elements like Mg, Ti, and likely others. The
argument is somewhat circular and cannot be easily solved, unless we
have additional information on the relative abundances of the enhanced
elements.

(v) What we learn from this systematic analysis is that the solution
is highly model dependent.  The results by TFWG00 and TMB03 can be
recovered only if the same pattern of abundances for the enhanced
elements is adopted, i.e. [Ti/Fe]$\sim$0 in particular. Passing to
SSPs with [Ti/Fe]$>$0 the tendency is to populate the old age bins of
the age distribution. We have taken as upper limit to the abundance
ratios ${\rm [X_{el}/Fe]}$ for some crucial elements like Mg and Ti
the values observed in Galactic Globular Clusters.

(vi) In addition to this, we have addressed the question whether the
solution found for age metallicity, and enhancement factor is
independent of the triplet of indices one has been using. The answer
is of course no, in the sense that different triplets having a
different resolving power leads to different results. Part of the
difficulty resides in the {\it Minimum-Distance Method} itself which
turns out to be inadequate to handle situations in which small
variations in the observational and/or theoretical indices imply large
variations in the age, metallicity and degree of enhancement. To cope
with this difficulty we tried to estimate ages, metallicities, and
enhancement in $\alpha$-elements by simultaneously fitting many
indices (six in our case). Though the situation gets better, it is not
fully satisfactory.

(viii) Needless to say that the results still depend on the input SSPs
and especially on the {\it Response Functions} in use. This is a point
of uncertainty that cannot be easily improved unless one has
independent assessments of the quality of stellar models and
calibrations. The {\it Response Functions}, in particular, are the
major drawback of the whole problem because they are so far available
only for three stars in total. A great deal of the results for the
$\alpha$-enhanced mixtures are still affected by the poor and coarse
calibration at disposal. \citet{Tantalo04c} are currently working on
this problem to provide {\it Response Functions} for a large grid of
effective temperatures and gravities over the whole HR-Diagram, a
large range of metallicities, and both solar and $\alpha$-enhanced
ratios. Some preliminary results of this study have already been
anticipated in Sect.~\ref{cal_trip}.
  
(ix) The sensitivity of indices like \Hbeta\ to some abundance
ratios (e.g. [Ti/Fe] and likely other elements) adds another dimension
to the interpretation of the two indices plane. We have indeed shown
how in presence of enhancement ($\Gamma >0$ and suitable ratios $\rm
[X_{el}/Fe]>0$, [Ti/Fe] in our case) an old galaxy could lie in the
same region of a young one with solar abundance ratios. Were this the
case, the observational spread in \Hbeta\ could be the signature of
different histories of star formation and chemical enrichment in turn
passing from one galaxy to another. This topic has been carefully
investigated in the companion paper by \citet{Tantalo04b} in which the
suggestion is advanced that part of the scatter along the \Hbeta\ axis
observed in the \Hbeta\ vs. \MgFe\ plane could be attributed instead
of the age, the current explanation, to a spread both in the degree of
enhancement and some abundance ratios.

(x) There is an important remark to be made about the use of SSPs to
simulate the complexity of a real galaxy. We have already touched upon
this topic in Sect.~\ref{mindist}, but it should be stressed once more
here. In real galaxies, even in the case of EGs, a mix of stellar
populations with different ages and chemical properties is likely to
exist. Therefore the approximation to SSPs is no longer valid and SSPs
should be replaced by galactic models incorporating the history of
star formation and chemical enrichment and the indices to be used
should take into account the contribution from all stellar
components. Some of the properties shown SSPs, especially those caused
by non standard evolutionary stages (e.g. reversal of indices like
\Hbeta\ at increasing age and metallicity), that now have the same
weight in the total balance of the resolving technique, are likely to
smear out in the complex mix of stars because the contribution by each
SSPs is proportional to the number of stars in the different age and
metallicity bins. Integrated indices for model galaxies have been
calculated by \citet{Tantalo98b} but never applied to this kind of
analysis. This is a point that should be carefully investigated.

(xi) According to the results of this analysis and of previous ones as
well, the conclusion one would draw is that EGs span very large ranges
of age (and also of metallicity and enhancement in
$\alpha$-elements). The age is, however, the most embarrassing
result. Does it means that in galaxies, for which formal ages younger
than the typical age of Globular Clusters (say 12\,Gyr) are found,
the bulk of stars have been formed at such young ages or there are
effects to be taken into account? The question has been addressed for
the first time by \citet{Bressan96}, explored in more detail by
\citet{Longhetti20}, and also discussed by TFWG00. The question is:
how much a recent, minute episode of star formation, engaging a small
fraction of the total mass, may alter the indices of an otherwise old
population of stars in EGs? The analyses was made by superposing to
the indices of an old population of stars the variation caused by a
recent episode of star formation. The result is that indices like
\Hbeta\ are strongly affected by even small percentages of young
stars: as long as star formation is active they jump to very high
values and when star formation is over they fall back to the original
value on a time scale of about 1\,Gyr. Other indices like \mgii,
\MFe\ are much less affected even if the companion chemical enrichment
may somewhat change them. In diagnostics planes like \Hbeta\ vs. \MFe\
the galaxy performs an extended loop elongated towards the \Hbeta\
axis, thus causing an artificial dispersion which could be interpreted
as an age dispersions, whereas what we really see is the transient
phase associated to the temporary stellar activity. The implications
of this have  carefully been investigated by \citet[][]{Tantalo04b}
with the aid of simulations of later star forming episodes of
different intensity and age superposed to an old population of stars.

\section*{Acknowledgements}
We would like to thank Roberto Caimmi, Laura Greggio, Ezio Pignatelli
and Lorenzo Piovan for many stimulating discussions. We also thank
Alberto Buzzoni for his very constructive comments as referee of the
paper.  This study has been financed by the Italian Ministry of
Education, University, and Research (MIUR), and the University of
Padua under the special contract ``Formation and evolution of
elliptical galaxies: the age problem''.

\bibliographystyle{mn2e}            
\bibliography{mnemonic,biblio}    

\begin{thebibliography}{}

\bibitem[\protect\citeauthoryear{Barbuy}{Barbuy}{1994}]{Barbuy94}
Barbuy B.,  1994, ApJ, 430, 218

\bibitem[\protect\citeauthoryear{Bertelli, Bressan, Chiosi, Fagotto \&
  Nasi}{Bertelli et~al.}{1994}]{Bertelli94}
Bertelli G.,  Bressan A.,  Chiosi C.,  Fagotto F.,    Nasi E.,  1994, A\&AS,
  106, 275

\bibitem[\protect\citeauthoryear{Borges, Idiart, de Freitas-Pacheco \&
  Thevein}{Borges et~al.}{1995}]{Borges95}
Borges C.~A.,  Idiart T.~P.,  de Freitas-Pacheco J.~A.,    Thevein F.,  1995,
  AJ, 110, 2408

\bibitem[\protect\citeauthoryear{Bressan, Chiosi \& Fagotto}{Bressan
  et~al.}{1994}]{Bressan94}
Bressan A.,  Chiosi C.,    Fagotto F.,  1994, ApJS, 94, 63

\bibitem[\protect\citeauthoryear{Bressan, Chiosi \& Tantalo}{Bressan
  et~al.}{1996}]{Bressan96}
Bressan A.,  Chiosi C.,    Tantalo R.,  1996, A\&A, 311, 425

\bibitem[\protect\citeauthoryear{Brocato, Matteucci, Mazzitelli \&
  Tornamb\'e}{Brocato et~al.}{1990}]{Brocato90}
Brocato E.,  Matteucci F.,  Mazzitelli I.,    Tornamb\'e A.,  1990, ApJ, 349,
  458

\bibitem[\protect\citeauthoryear{Burstein, Bertola, Buson, Faber \&
  Lauer}{Burstein et~al.}{1988}]{Burstein88}
Burstein D.,  Bertola F.,  Buson L.~M.,  Faber S.~M.,    Lauer T.~R.,  1988,
  ApJ, 328, 440

\bibitem[\protect\citeauthoryear{Burstein, Faber, Gaskell \& Krumm}{Burstein
  et~al.}{1984}]{Burstein84}
Burstein D.,  Faber S.~M.,  Gaskell C.~M.,    Krumm N.,  1984, ApJ, 287, 586

\bibitem[\protect\citeauthoryear{Carney}{Carney}{1996}]{Carney96}
Carney B.,  1996, PASP, 108, 900

\bibitem[\protect\citeauthoryear{Castellani \& Tornamb\'e}{Castellani \&
  Tornamb\'e}{1991}]{CasTor91}
Castellani V.,  Tornamb\'e A.,  1991, apj, 381, 393

\bibitem[\protect\citeauthoryear{Chiosi, Bertelli \& Bressan}{Chiosi
  et~al.}{1992}]{Chiosi92}
Chiosi C.,  Bertelli G.,    Bressan A.,  1992, ARA\&A, 30, 235

\bibitem[\protect\citeauthoryear{Chiosi \& Carraro}{Chiosi \&
  Carraro}{2002}]{Chiocar02}
Chiosi C.,  Carraro G.,  2002, MNRAS, 335, 335

\bibitem[\protect\citeauthoryear{Davies, Kuntschner, Emsellem, Bacon, Bureau,
  Carollo, Copin, Miller, Monnet, Peletier, Verolme \& de Zeeuw}{Davies
  et~al.}{2001}]{Davies2001}
Davies R.~L.,  Kuntschner H.,  Emsellem E.,  Bacon R.,  Bureau M.,  Carollo
  C.~M.,  Copin Y.,  Miller B.~W.,  Monnet G.,  Peletier R.~F.,  Verolme E.~K.,
     de Zeeuw P.~T.,  2001, ApJL, 548, L33

\bibitem[\protect\citeauthoryear{Dorman, Rood \& O'Connell}{Dorman
  et~al.}{1993}]{Dorman93}
Dorman B.,  Rood R.~T.,    O'Connell R.~W.,  1993, A\&A, 419, 516

\bibitem[\protect\citeauthoryear{Faber, Friel, Burstein \& Gaskell}{Faber
  et~al.}{1985}]{Faber85}
Faber S.~M.,  Friel E.~D.,  Burstein D.,    Gaskell C.~M.,  1985, ApJS, 57, 711

\bibitem[\protect\citeauthoryear{Faber, Worthey \& Gonz\'alez}{Faber
  et~al.}{1992}]{Faber92}
Faber S.~M.,  Worthey G.,    Gonz\'alez J.~J.,  1992, in Barbuy B.,  Renzini
  A.,  eds, {\it The Stellar Population of Galaxies} IAU Symp. 149.
Kluwer Academic Publishers: Dordrecht, p.~255

\bibitem[\protect\citeauthoryear{Fagotto, Bressan, Bertelli \& Chiosi}{Fagotto
  et~al.}{1994}]{Fagotto94c}
Fagotto F.,  Bressan A.,  Bertelli G.,    Chiosi C.,  1994, A\&AS, 105, 39

\bibitem[\protect\citeauthoryear{Girardi \& Bertelli}{Girardi \&
  Bertelli}{1998}]{GirBer98}
Girardi L.,  Bertelli G.,  1998, MNRAS, 300, 533

\bibitem[\protect\citeauthoryear{Girardi, Bressan, Bertelli \& Chiosi}{Girardi
  et~al.}{2000}]{Girardi20}
Girardi L.,  Bressan A.,  Bertelli G.,    Chiosi C.,  2000, A\&AS, 141, 371

\bibitem[\protect\citeauthoryear{Girardi, Bressan, Chiosi, Bertelli \&
  Nasi}{Girardi et~al.}{1996}]{Girardi96}
Girardi L.,  Bressan A.,  Chiosi C.,  Bertelli G.,    Nasi E.,  1996, A\&A,
  117, 113

\bibitem[\protect\citeauthoryear{Gonz\'alez}{Gonz\'alez}{1993}]{Gonzalez93}
Gonz\'alez J.~J.,  1993, PhD thesis, University of California, Santa Cruz

\bibitem[\protect\citeauthoryear{Gratton, Carretta, Claudi, Lucatello \&
  Barbieri}{Gratton et~al.}{2003}]{Gratton03}
Gratton R.,  Carretta E.,  Claudi R.,  Lucatello S.,    Barbieri M.,  2003,
  A\&A, 404, 187

\bibitem[\protect\citeauthoryear{Green, Demarque \& King}{Green
  et~al.}{1987}]{Green87}
Green E.,  Demarque P.,    King C.,  1987, The revised Yale isochrones and
  luminosity functions.
New Haven: Yale Observatory, 1987

\bibitem[\protect\citeauthoryear{Greggio \& Renzini}{Greggio \&
  Renzini}{1983}]{Greggio83}
Greggio L.,  Renzini A.,  1983, A\&A, 118, 217

\bibitem[\protect\citeauthoryear{Greggio \& Renzini}{Greggio \&
  Renzini}{1990}]{Greggio90}
Greggio L.,  Renzini A.,  1990, ApJ, 364, 599

\bibitem[\protect\citeauthoryear{Grevesse, Noels \& Sauval}{Grevesse
  et~al.}{1996}]{Grevesse96}
Grevesse N.,  Noels A.,    Sauval A.,  1996, in Holt S.,  Sonneborn G.,  eds,
  {\it Cosmic Abundances} ASP Conf. Ser. 99.
San Francisco: ASP, p.~117

\bibitem[\protect\citeauthoryear{Habgood}{Habgood}{2001}]{Habgood01}
Habgood M.-J.,  2001, PhD thesis, Univ. of North Carolina at Chapel Hill

\bibitem[\protect\citeauthoryear{Horch, Demarque \& Pinsonneault}{Horch
  et~al.}{1992}]{Horch92}
Horch E.,  Demarque P.,    Pinsonneault M.,  1992, ApJ, 388, L53

\bibitem[\protect\citeauthoryear{Idiart \& de Freitas-Pacheco}{Idiart \&
  de~Freitas-Pacheco}{1995}]{Idiart95}
Idiart T.~P.,  de Freitas-Pacheco J.~A.,  1995, AJ, 109, 2218

\bibitem[\protect\citeauthoryear{J{\o}rgensen}{J{\o}rgensen}{1999}]{Jorgensen9%
9}
J{\o}rgensen I.,  1999, MNRAS, 306, 607

\bibitem[\protect\citeauthoryear{Kuntschner}{Kuntschner}{1998}]{Kuntschner98}
Kuntschner H.,  1998, PhD thesis, Univ. of Durham

\bibitem[\protect\citeauthoryear{Kuntschner}{Kuntschner}{2000}]{Kuntschner00}
Kuntschner H.,  2000, MNRAS, 315, 184

\bibitem[\protect\citeauthoryear{{Kuntschner} \& {Davies}}{{Kuntschner} \&
  {Davies}}{1998}]{Kuntschner98a}
{Kuntschner} H.,  {Davies} R.~L.,  1998, MNRAS, 295, L29

\bibitem[\protect\citeauthoryear{Kuntschner, Lucey, Smith, Hudsen \&
  Davies}{Kuntschner et~al.}{2001}]{Kuntschner01b}
Kuntschner H.,  Lucey J.~R.,  Smith R.~J.,  Hudsen M.~J.,    Davies R.~L.,
  2001, MNRAS, 323, 615

\bibitem[\protect\citeauthoryear{Larson}{Larson}{1974}]{Larson74}
Larson R.~B.,  1974, MNRAS, 142, 501

\bibitem[\protect\citeauthoryear{Leitherer, Alloin, v. Alvensleben, Gallagher,
  Huchra \& et al.}{Leitherer et~al.}{1996}]{Leitherer96}
Leitherer C.,  Alloin D.,  v. Alvensleben U.~F.,  Gallagher J.,  Huchra J.,
  et al. 1996, PASP, 108, 996

\bibitem[\protect\citeauthoryear{{Longhetti}, {Bressan}, {Chiosi} \&
  {Rampazzo}}{{Longhetti} et~al.}{2000}]{Longhetti20}
{Longhetti} M.,  {Bressan} A.,  {Chiosi} C.,    {Rampazzo} R.,  2000, A\&A,
  353, 917

\bibitem[\protect\citeauthoryear{Maraston}{Maraston}{1998}]{Maraston98}
Maraston C.,  1998, MNRAS, 300, 872

\bibitem[\protect\citeauthoryear{Maraston, Greggio, Renzini, S.Ortolani,
  Saglia, Puzia \& Kissler-Patig}{Maraston et~al.}{2003}]{Maraston03}
Maraston C.,  Greggio L.,  Renzini A.,  S.Ortolani Saglia R.,  Puzia T.,
  Kissler-Patig M.,  2003, A\&A, 400, 823

\bibitem[\protect\citeauthoryear{Marigo, Chiosi \& Kudritzki}{Marigo
  et~al.}{2002}]{MarChiKud02}
Marigo P.,  Chiosi C.,    Kudritzki R.-P.,  2002, A\&A, p.~617

\bibitem[\protect\citeauthoryear{Matteucci}{Matteucci}{1994}]{Mattfra94}
Matteucci F.,  1994, A\&A, 154, 279

\bibitem[\protect\citeauthoryear{Matteucci}{Matteucci}{1997}]{Matteucci97}
Matteucci F.,  1997, Fundam. Cosmic Phys., 17, 283

\bibitem[\protect\citeauthoryear{Matteucci, Ponzone \& Gibson}{Matteucci
  et~al.}{1998}]{Matteucci98}
Matteucci F.,  Ponzone R.,    Gibson B.~K.,  1998, A\&A, 335, 855

\bibitem[\protect\citeauthoryear{Munari, Sordo, Castelli \& Zwitter}{Munari
  et~al.}{2004}]{Munari04}
Munari U.,  Sordo R.,  Castelli F.,    Zwitter T.,  2004, A\&A, to be submitted

\bibitem[\protect\citeauthoryear{Pagel}{Pagel}{1989}]{Pagel89}
Pagel B. E.~J.,  1989, in Beckman J.,  Page B. E.~J.,  eds, {\it Evolutionary
  Phenomena in Galaxies} Cambridge University Press: Cambridge, p.~201

\bibitem[\protect\citeauthoryear{Peimbert \& Peimbert}{Peimbert \&
  Peimbert}{2002}]{Peimbert02}
Peimbert A.,  Peimbert M.,  2002, in Revista Mexicana de Astronomia y
  Astrofisica Conference Series {\it A New Determination of the Primordial
  Helium Abundance}.
pp 250--250

\bibitem[\protect\citeauthoryear{Poggianti, Bridges, Mobasher, Carter, Doi,
  Iye, Kashikawa, Komiyama, Okamura, Sekiguchi, Shimasaku, Yagi \&
  Yasuda}{Poggianti et~al.}{2001}]{Poggianti01}
Poggianti B.,  Bridges T.,  Mobasher B.,  Carter D.,  Doi M.,  Iye M.,
  Kashikawa N.,  Komiyama Y.,  Okamura S.,  Sekiguchi M.,  Shimasaku K.,  Yagi
  M.,    Yasuda N.,  2001, ApJ, 562, 689

\bibitem[\protect\citeauthoryear{Reimers}{Reimers}{1975}]{Reimers75}
Reimers D.,  1975, Mem. Soc. R. Sci. Liege, ser. 6, 8, 369

\bibitem[\protect\citeauthoryear{Renzini \& Buzzoni}{Renzini \&
  Buzzoni}{1986}]{Renbuz86}
Renzini A.,  Buzzoni A.,  1986, in Chiosi C.,  Renzini A.,  eds, {\it Spectral
  Evolution of Galaxies} Dordecht: Riedel, p.~213

\bibitem[\protect\citeauthoryear{Ryan, Norris \& Bessell}{Ryan
  et~al.}{1991}]{Ryan91}
Ryan S.,  Norris J.,    Bessell M.,  1991, AJ, 102, 303

\bibitem[\protect\citeauthoryear{Salasnich, Girardi, Weiss \& Chiosi}{Salasnich
  et~al.}{2000}]{Salasnich20}
Salasnich B.,  Girardi L.,  Weiss A.,    Chiosi C.,  2000, A\&A, 361, 1023, (SGWC00)

\bibitem[\protect\citeauthoryear{Salvaterra \& Ferrara}{Salvaterra \&
  Ferrara}{2003}]{Salvaterra03}
Salvaterra R.,  Ferrara A.,  2003, MNRAS, 340, L17

\bibitem[\protect\citeauthoryear{Tantalo}{Tantalo}{1998}]{Tantalo98}
Tantalo R.,  1998, PhD thesis, Univ. of Padova

\bibitem[\protect\citeauthoryear{Tantalo \& Chiosi}{Tantalo \&
  Chiosi}{2004a}]{Tantalo04d}
Tantalo R.,  Chiosi C.,  2004a, MNRAS, in preparation

\bibitem[\protect\citeauthoryear{Tantalo \& Chiosi}{Tantalo \&
  Chiosi}{2004b}]{Tantalo04b}
Tantalo R.,  Chiosi C.,  2004b, MNRAS, accepted

\bibitem[\protect\citeauthoryear{Tantalo, Chiosi \& Bressan}{Tantalo
  et~al.}{1998}]{Tantalo98a}
Tantalo R.,  Chiosi C.,    Bressan A.,  1998, A\&A, 333, 419

\bibitem[\protect\citeauthoryear{Tantalo, Chiosi, Bressan, Marigo \&
  Portinari}{Tantalo et~al.}{1998}]{Tantalo98b}
Tantalo R.,  Chiosi C.,  Bressan A.,  Marigo P., Portinari L.,  1998, A\&A,
  335, 823

\bibitem[\protect\citeauthoryear{Tantalo, Chiosi, Munari, Piovan \&
  Sordo}{Tantalo et~al.}{2004}]{Tantalo04c}
Tantalo R.,  Chiosi C.,  Munari U.,  Piovan L., Sordo R., 2004, MNRAS, submitted

\bibitem[\protect\citeauthoryear{Thomas \& Maraston}{Thomas \&
  Maraston}{2003}]{ThoMara03}
Thomas D.,  Maraston C.,  2003, A\&A, 401, 429

\bibitem[\protect\citeauthoryear{Thomas, Maraston \& Bender}{Thomas
  et~al.}{2003a}]{Thomas03}
Thomas D.,  Maraston C., Bender R.,  2003a, MNRAS, 339, 897, (TMB03)

\bibitem[\protect\citeauthoryear{Thomas, Maraston \& Bender}{Thomas
  et~al.}{2003b}]{Thomas03a}
Thomas D.,  Maraston C.,    Bender R.,  2003b, MNRAS, 343, 279

\bibitem[\protect\citeauthoryear{Trager}{Trager}{1997}]{Trager97}
Trager S.,  1997, PhD thesis, Univ. of California, Santa Cruz

\bibitem[\protect\citeauthoryear{Trager, Faber, Worthey \& Gonz\'alez}{Trager
  et~al.}{2000a}]{Trager20b}
Trager S.~C.,  Faber S.~M.,  Worthey G., Gonz\'alez J.~J.,  2000a, AJ, 120,
  165

\bibitem[\protect\citeauthoryear{Trager, Faber, Worthey \& Gonz\'alez}{Trager
  et~al.}{2000b}]{Trager20a}
Trager S.~C.,  Faber S.~M.,  Worthey G.,    Gonz\'alez J.~J.,  2000b, AJ, 119,
  1645, (TFWG00)

\bibitem[\protect\citeauthoryear{Tripicco \& Bell}{Tripicco \&
  Bell}{1995}]{Tripicco95}
Tripicco M.~J.,  Bell R.~A.,  1995, AJ, 110, 3035, (TB95)

\bibitem[\protect\citeauthoryear{Vandenberg}{Vandenberg}{1985}]{Vandenberg85}
Vandenberg D.,  1985, ApJS, 58, 711

\bibitem[\protect\citeauthoryear{Vandenberg \& Bell}{Vandenberg \&
  Bell}{1985}]{VandenbergBell85}
Vandenberg D.,  Bell R.,  1985, ApJS, 58, 561

\bibitem[\protect\citeauthoryear{Vandenberg \& Laskarides}{Vandenberg \&
  Laskarides}{1987}]{Vanderberg87}
Vandenberg D.,  Laskarides P.,  1987, ApJS, 64, 103

\bibitem[\protect\citeauthoryear{Vassiliadis \& Wood}{Vassiliadis \&
  Wood}{1993}]{Vassiliadis93}
Vassiliadis D.~A.,  Wood P.~R.,  1993, ApJ, 413, 641

\bibitem[\protect\citeauthoryear{Vazdekis, Kuntschner, Davies, Arimoto,
  Nakamura \& Peletier}{Vazdekis et~al.}{2001}]{Vazdekis01}
Vazdekis A.,  Kuntschner H.,  Davies R.~L.,  Arimoto N.,  Nakamura O.,
  Peletier R.~F.,  2001, apj, 551, 127

\bibitem[\protect\citeauthoryear{Weiss, Peletier \& Matteucci}{Weiss
  et~al.}{1995}]{Weiss95}
Weiss A.,  Peletier R.~F.,    Matteucci F.,  1995, A\&A, 296, 73

\bibitem[\protect\citeauthoryear{Worthey}{Worthey}{1992}]{Worthey92}
Worthey G.,  1992, PhD thesis, Univ. of California

\bibitem[\protect\citeauthoryear{Worthey}{Worthey}{1994}]{Worthey94a}
Worthey G.,  1994, ApJS, 95, 107

\bibitem[\protect\citeauthoryear{Worthey, Faber \& Gonz\'alez}{Worthey
  et~al.}{1992}]{Worthey92b}
Worthey G.,  Faber S.~M.,    Gonz\'alez J.~J.,  1992, ApJ, 398, 69

\bibitem[\protect\citeauthoryear{Worthey, Faber, Gonz\'alez \&
  Burstein}{Worthey et~al.}{1994}]{Worthey94}
Worthey G.,  Faber S.~M.,  Gonz\'alez J.~J.,    Burstein D.,  1994, ApJS, 94,
  687

\bibitem[\protect\citeauthoryear{Worthey \& Ottaviani}{Worthey \&
  Ottaviani}{1997}]{Ottaviani97}
Worthey G.,  Ottaviani D.,  1997, ApJS, 111, 377

\end{thebibliography}

\end{document}